\documentclass[journal]{IEEEtran}

\usepackage[caption=false]{subfig}
\usepackage{cite}
\usepackage{amsmath}    	% further math commands
\usepackage{amssymb}	    % further math symbols
\usepackage{amsfonts}		% further math symbols
\usepackage{amsthm}    	    % theorems
\usepackage{mathtools}		% further math commands
\usepackage{bm}             % for greek bold letters
\usepackage{cases}
\usepackage[ruled,vlined,linesnumbered]{algorithm2e}
\usepackage{tikz}
\usepackage{pgfplots}
\usepackage{stackengine}
\usepackage{soul}
\usepackage{multirow}
\usepackage[normalem]{ulem}
\usepackage{siunitx}
\usepackage{steinmetz}
\usepackage[shortlabels]{enumitem}
\usepackage{forest}
\usepackage{accents}
\usepackage[export]{adjustbox}  % to use max width in includegraphics
\usepackage{lipsum}
\usepackage{hyperref}
\usepackage{url}
\usepackage{etoolbox}
\usepackage{booktabs}   	      % improve tables

\newcommand{\T}{^{\mathsf{T}}}
\newcommand{\B}[1]{\if#1\relax\bm{#1}\else\mathbf{#1}\fi} % bold
\newcommand{\R}[1]{\mathrm{#1}}  % regul. text
\newcommand{\C}[1]{\mathcal{#1}}
\newcommand{\BB}[1]{\mathbb{#1}}

\newcommand{\norm}[1]{\left\lVert #1 \right\rVert}
\newcommand{\abs}[1]{\left\lvert #1 \right\rvert}
\newcommand{\wrapalgo}{\quad\textcolor{gray}{\(\hookrightarrow\)}\quad}

\SetKwComment{tcc}{ {$\triangleright$} }{} % comment C style
\SetKwComment{tcp}{ {$\triangleright$} }{} % comment C++ style

\SetCommentSty{myCommentStyle}
\makeatletter
\patchcmd{\@algocf@start}% <cmd>
  {-1.5em}% <search>
  {0pt}% <replace>
  {}{}% <success><failure>
\makeatother

\newtheorem{theorem}{Theorem}[section]
\newtheorem{corollary}[theorem]{Corollary}
\newtheorem{proposition}[theorem]{Proposition}
\newtheorem{lemma}[theorem]{Lemma}
\newtheorem{definition}[theorem]{Definition}
\newtheorem{remark}[theorem]{Remark}
\newtheorem{assumption}[theorem]{Assumption}
\newtheorem{problem}[theorem]{Problem}
 
\tikzstyle{block} = [draw, rectangle, 
    minimum height=3em, minimum width=5em]
\tikzstyle{input} = [coordinate]
\tikzstyle{output} = [coordinate]
\tikzstyle{pinstyle} = [pin edge={to-,thin,black}]
\usetikzlibrary{shapes.geometric,shapes.arrows,decorations.pathmorphing,calc}
\usetikzlibrary{matrix,chains,scopes,positioning,arrows,fit}

\allowdisplaybreaks

\pgfplotsset{compat=1.18}

\begin{document}

\title{Guaranteeing Control Requirements\\ via Reward Shaping in Reinforcement Learning\\}

\author{Francesco De Lellis~\IEEEmembership{Member,~IEEE}, Marco Coraggio~\IEEEmembership{Member,~IEEE}, Giovanni Russo~\IEEEmembership{Senior Member,~IEEE}, Mirco Musolesi~\IEEEmembership{Member,~IEEE}, Mario di Bernardo~\IEEEmembership{Fellow,~IEEE}
\thanks{
This work was in part supported by the Research Project “SHARESPACE”
funded by the European Union (EU HORIZON-CL4-2022-
HUMAN-01-14. SHARESPACE. GA 101092889 - http://sharespace.eu) and by {\color{black}the Research Project PRIN 2022 “Machine-learning based control of complex multi-agent systems for search and rescue operations in natural disasters (MENTOR)”
funded by the Italian Ministry of University and Research (2023--2025).}

F. De Lellis is with the Department of Electrical Engineering and Information Technology, University of Naples Federico II, Naples, Italy (e-mail: francesco.delellis@unina.it).

M. Coraggio is with the Scuola Superiore Meridionale, School for Advanced Studies, Naples, Italy (e-mail: marco.coraggio@unina.it).

G. Russo is with the Department of Computer and Electrical Engineering \& Applied Mathematics, University of Salerno, DIEM,  Salerno, Italy (e-mail: giovarusso@unisa.it).

M. Musolesi is with the Department of Computer Science, University College London, U.K. and the Department of Informatics - Science and Engineering, University of Bologna, Bologna, Italy (e-mail: m.musolesi@ucl.ac.uk).

M. di Bernardo is with the Department of Electrical Engineering and Information Technology, University of Naples Federico II, Naples, Italy,
and with the Scuola Superiore Meridionale, School for Advanced Studies, Naples, Italy (e-mail: mario.dibernardo@unina.it).}%
}%

% \markboth{Journal of \LaTeX\ Class Files,~Vol.~XXX, No.~XXX, July~2023}%
% {Author \MakeLowercase{\textit{et al.}}: TITLE}

% Remember, if you use this you must call \IEEEpubidadjcol in the second column for its text to clear the IEEEpubid mark.

\maketitle

% abstract and keywords -----------------------
\begin{abstract}
In addressing control problems such as regulation and tracking through reinforcement learning, it is often required to guarantee that the acquired policy meets essential performance and stability criteria such as a desired settling time and steady-state error prior to deployment. Motivated by this necessity, we present a set of results and a systematic reward shaping procedure 
that (i) ensures the optimal policy generates trajectories that align with specified control requirements and (ii) allows to assess whether any given policy satisfies them.
%designed to assess whether a policy can generate trajectories aligning with specified control requirements. 
We validate our approach through comprehensive numerical experiments conducted in two representative environments from OpenAI Gym: the Inverted Pendulum swing-up problem and the Lunar Lander. Utilizing both tabular and deep reinforcement learning methods, our experiments consistently affirm the efficacy of our proposed framework, highlighting its effectiveness in ensuring policy adherence to the prescribed control requirements.
\end{abstract}

\begin{IEEEkeywords}
Learning-Based Control, Reward Shaping, Policy validation, Deep Reinforcement Learning, Computational Control
\end{IEEEkeywords}

%--------------------------------
\section{Introduction}
\label{sec:introduction}

% \subsection{{\color{black}Problem description and motivation}}

The paradigm of using reinforcement learning (RL) for control system design has gained substantial traction due to its ability to autonomously learn policies that effectively address complex control problems, relying solely on data and employing a reward maximization process. This approach finds diverse applications, spanning from attitude control \cite{dong2021reinforcement} and wind farm management \cite{dong2023datadriven} to autonomous car-driving \cite{9351818} and the regulation of plasma using high-fidelity simulators \cite{degrave2022magnetic}. However, a significant challenge in this domain revolves around ensuring that the learned control policy demonstrates the desired closed-loop performance and steady-state error, posing a crucial open question in control system design.

It is often argued that accurate knowledge of system dynamics is necessary to provide analytical guarantees of stability and performance, which is crucial for industrial applications \cite{5227780,osinenko2022reinforcement}.
In fact, in this paper, we introduce a set of analytical results and a constructive procedure for shaping the reward function of approaches based on reinforcement learning (tabular and function approximation methods that rely on deep learning).
The goal is to derive a learned policy which is obtained without the use of a mathematical model of the system dynamics, able to verifiably meet predetermined control requirements in terms of {\color{black}desired} settling time and {\color{black}steady-state} error.

\IEEEpubidadjcol

{\color{black}In the Literature, \emph{reward shaping}, consisting of modifying the reward function in order to improve learning or control performance, has mostly been used to increase sample efficiency \cite{randlov1998learning,ng1999policy,gupta2022unpacking}, rather than provide guarantees on the learned policy.
An early example was presented in \cite{randlov1998learning}, where an agent was trained to ride a bicycle exploiting a reward shaping mechanism.
More recently, reduced sample complexity was demonstrated in \cite{gupta2022unpacking} for a modified Upper Confidence Bound algorithm using shaped rewards.
In \cite{ng1999policy}, it was shown that adding a function of the state to the reward keeps the optimal policy unchanged if and only if the function is potential-based.
A method to select potential-based functions is presented and validated analytically in \cite{dong2020principleda}, requiring knowledge of an appropriate Lyapunov function, to guarantee convergence to a state under the optimal policy.
While this result can be used to solve regulation problems, it does not ensure a specific settling time. Moreover, finding a Lyapunov function is often cumbersome for many real-world problems.}

{\color{black}When guarantees are given on reinforcement learning control \cite{osinenko2022reinforcement}, they are typically provided in terms of reachability of certain subsets of the state space \cite{perkins2002lyapunov,berkenkamp2017safe}, or in terms of \emph{safety} during learning and/or for the learned policy.
Namely, in \cite{perkins2002lyapunov}, RL is used to select a control law among a set of candidates, using Lyapunov functions to ensure a system enters a goal region with unitary probability, under certain conditions on the controllers. 
In \cite{berkenkamp2017safe}, a partially known system model is used to improve a safe starting policy through RL, avoiding actions that bring the system out of the basin of attraction of a desired equilibrium.
Both approaches \cite{perkins2002lyapunov,berkenkamp2017safe} do not provide guarantees on the time required to reach the desired regions.
Safety for RL control has been extensively explored in the Literature using various frameworks, such as \emph{constrained Markov decision processes} \cite{chow2018lyapunovbased},
``shields'' \cite{alshiekh2018safe}, 
\emph{control barrier functions} \cite{marvi2021safe}, 
and a combination of Model Predictive Control and RL \cite{beckenbach2020qlearning}. 
Although these techniques ensure avoidance of unsafe subsets of the state space, they generally do not provide guarantees on reaching a specific goal region or on control performance metrics, such as settling time.
}

{\color{black}
The problem of synthesizing rewards for control tasks is also the subject of \emph{inverse optimal control} (IOC) \cite{10.1115/1.3653115}, focusing on estimating the rewards associated to given observations of states and control inputs, assuming closed-loop stability and/or policy optimality.
Initially aimed at determining control functions producing observed outputs \cite{506395}, IOC has since been connected to reinforcement learning \cite{Rodrigues}, applied in nonlinear, stochastic environments \cite{garrabe2023inverse}, and its framework has been used to investigate the cost design problem \cite{9429728}.
However, to the best of our knowledge, IOC has not been used specifically to design reward functions that, when optimized for, can guarantee specific control performance.
}

Given a regulation or tracking problem with predetermined stability and performance requirements on {\color{black}steady-state} error and settling time,
{\color{black}we advance the state of the art} as follows.
\begin{enumerate}
    \item We introduce a model-free sufficient condition on the discounted return associated to a trajectory to determine if {\color{black}it} is \emph{acceptable} ({\color{black}i.e.,} it satisfies the control requirements).
    \item We {\color{black}give} a sufficient condition to assess whether a learned policy leads to an acceptable closed-loop trajectory.
    \item We {\color{black}provide} a procedure to shape the reward function so that the above conditions can be applied on a system of interest and that the optimal policy is acceptable.
    \item We {\color{black}successfully} validate the approach {\color{black}through} two representative control problems from OpenAI Gym \cite{brockman2016openai}: the stabilization of an inverted pendulum \cite{GYM_pendulum}, and performing landing in the Lunar Lander environment \cite{GYM_lander}. 
    % All experiments confirm the effectiveness of our approach.
\end{enumerate}

For reproducibility, the code is available on GitHub \cite{code}.

The rest of the paper is organized as follows.
In Section \ref{sec:problem_statement}{\color{black},} we formalize the problem of constructing a reward function for learning-based control.
The main results of our approach are then presented in Section \ref{sec:main_result} and validated via numerical simulations on two representative application examples in Section \ref{sec:numerical_results}.
Concluding remarks are given in Section \ref{sec:conclusions}.

%--------------------------------
\section{Problem statement}
\label{sec:problem_statement}

%--------------------------------------------
\subsection{Problem set-up}
\label{sec:problem_setup}

We consider a discrete time dynamical system of the form
\begin{equation}\label{eq:dynamical_system}
    x_{k+1} = f(x_k, u_k),
    \quad x_0 = \tilde{x}_0,
\end{equation}
where $k \in \BB{N}_{\ge 0}$ is discrete time,
$x_k \in \C{X}$ is the \emph{state} at time $k$, 
$\C{X}$ is the \emph{state space},
$\tilde{x}_0 \in \C{X}$ is an \emph{initial condition}, 
$u_k \in \C{U}$ is the \emph{control input} (or \emph{action}) at time $k$, 
$\C{U}$ is the \emph{set of feasible inputs}, 
and $f : \C{X} \times \C{U} \rightarrow \C{X}$ is the system \emph{dynamics}.

Furthermore, we let $\pi : \C{X} \rightarrow \C{U}$ be a \emph{control policy}, and let $u_k = \pi(x_k)$.
Let also $\C{X}^\infty \coloneqq \C{X} \times \cdots \times \C{X}$, with the Cartesian product being applied an infinite number of times.
We denote by $\phi^{\pi}(\tilde{x}_0) \in \C{X}^\infty$ the \emph{trajectory} obtained by applying policy $\pi$ to system \eqref{eq:dynamical_system} starting from $\tilde{x}_0$ as initial state.

We are interested in finding a policy such that the trajectory generated by it (starting from a given $\tilde{x}_0$) reaches a desired \emph{goal region} $\C{G} \subset \C{X}$ before some \emph{{\color{black}desired} settling time} $k_\R{s} \in \BB{N}_{>0}$ and remains in this region for at least a \emph{{\color{black}desired} permanence time} $k_\R{p} \in \BB{N}_{>0}$ {\color{black}(see Definition \ref{def:acceptable} below for the rigorous statements)}.
{\color{black}For example, $\C{G}$} could be an arbitrarily small neighborhood of a reference state, {\color{black}with a radius equal to the admitted steady-state error}.
{\color{black}In our main results, we assume $\C{G}$, $k_\R{s}$, $k_\R{p}$ are given; nonetheless,} in Section \ref{sec:main_result} (see Remark \ref{rem:selection_permanence_time}), we will {\color{black}observe} that $k_\R{p}$ can be arbitrarily large, {\color{black}and in Proposition \ref{pro:settling_time_large_enough} (in the Appendix), we give a criterion 
to assess the feasibility of the settling time constraint when} limited knowledge about the system to control is available.

%--------------------------------------------
\subsection{Acceptable state{\color{black}-space} sequences}
\label{sec:acceptable_state_sequences}

We will now introduce concepts that will be used for the formalization of the proposed approach.
\begin{definition}[{\color{black}State-space sequences}]\label{def:state_space_sequences}
    A \emph{{\color{black}state-space sequence}} is a sequence $\xi = (x_k)_{k \in [0, +\infty)} \in \C{X}^\infty$.
\end{definition}

\begin{figure}[t]
  \centering
  \subfloat[]{\includegraphics{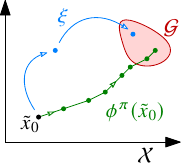}
  \label{fig:state_space_sequence_trajectory}}%
  \qquad
  \subfloat[]{\includegraphics{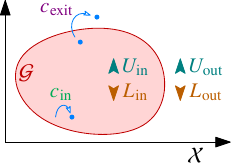}
  \label{fig:reward_structure}}
  \caption{(a): A {\color{black}state-space sequence} $\xi$, a trajectory $\phi^\pi(\tilde{x}_0)$, and a goal region $\C{G}$ (see {\color{black}Section} \ref{sec:problem_statement}); while a {\color{black}state-space sequence} is simply a sequence of points in the state space $\C{X}$, a trajectory is generated by applying a policy to the dynamics
  %described by
  {\color{black}in} \eqref{eq:dynamical_system}.
  (b): Terms of the reward structure in Assumption \ref{ass:reward_structure}.%
  }
\end{figure}

Note that all trajectories are {\color{black}state-space sequences}, but the converse is not true.
As a matter of fact, given a {\color{black}state-space sequence} $\xi$ with $x_0 = \tilde{x}_0 \in \C{X}$, there is no guarantee that there exists a policy $\pi$ such that $\phi^{\pi}(\tilde{x}_0) = \xi$.
A graphical representation of these concepts is reported in Figure \ref{fig:state_space_sequence_trajectory}.

\begin{definition}[First exit instant]\label{def:first_exit_instant}
    The \emph{first exit instant} $k_{\R{exit}}(\xi) \in \BB{N}_{> 0}$ of a {\color{black}state-space sequence} $\xi = (x_k)_{k \in [0, +\infty)}$ is the smallest time instant such that, in $\xi$, we have $x_{k_{\R{exit}}(\xi) - 1} \in \C{G}$ and $x_{k_{\R{exit}}(\xi)} \not\in \C{G}$; if this condition never occurs in $\xi$, we set $k_{\R{exit}}(\xi) = \infty$.    
\end{definition}

Next, we define the set of acceptable {\color{black}state-space sequences}, trajectories, and policies, i.e., those that satisfy the performance and {\color{black}steady-state} specifications.
%comply with our aim.

\begin{definition}[Acceptable {\color{black}state-space sequences}, trajectories and policies]%
\label{def:acceptable}
Given the desired goal region $\C{G}$, the {\color{black}desired} settling time $k_\R{s}$, and the {\color{black}desired} permanence time $k_\R{p}$, a {\color{black}state-space sequence} $\xi = (x_0, x_1, x_2, \dots)$ or equivalently a trajectory $\phi^\pi(\tilde{x}_0) = (\tilde{x}_0, x_1, x_2, \dots)$ are 
%said to be 
\emph{acceptable} if
\begin{enumerate}
    \item\label{ite:constraint_settling_time}
    $\exists \, k \le k_{\R{s}} : x_k \in \C{G}$ 
    (i.e., the state is in $\C{G}$ not later than time $k_{\R{s}}$);
    \item\label{ite:constraint_permanence_time}
    $k_{\R{exit}}(\xi) > k_{\R{p}}$ 
    (i.e., the state does not exit $\C{G}$ before time $k_{\R{p}}$, included).
\end{enumerate}
A policy $\pi$ is \emph{acceptable from $\tilde{x}_0$} if $\phi^\pi(\tilde{x}_0)$ is acceptable.
\end{definition}
It {\color{black}can be immediately verified} that there exists at least one acceptable {\color{black}state-space sequence} provided that $\C{G} \ne \varnothing$.
Indeed, this {\color{black}state-space sequence} is $\xi = (x_0, x_1, \dots)$ with $x_k \in \C{G}$ for all $k$, which can be verified to be acceptable by checking the two conditions in Definition~\ref{def:acceptable}.

%------------------------------------------
\subsection{Using reinforcement learning to find acceptable control policies}
\label{sec:using_rl_to_find_acceptable_policies}

Following \cite{pmlr-v211-de-lellis23a, pmlr-v168-lellis22a}, we employ a reinforcement learning solution to automatically identify an acceptable policy for a given initial condition $\tilde{x}_0$, and to do so without the need of knowing the dynamics $f$.
Namely, let $r : \C{X} \times \C{X} \times \C{U} \rightarrow \BB{R}$ be a \emph{reward function}, so that $r(x', x, u)$ is the reward obtained by the agent when taking action $u$ in state $x$ and arriving at the new state $x'$ at the next time instant.
Let also $J^\pi : \C{X}^\infty \rightarrow \BB{R}$ be the \emph{(discounted) return function} defined as
\begin{equation}\label{eq:return}
    J^\pi(\xi) \coloneqq \sum_{k=1}^{\infty} \gamma^{k-1} r(x_{k}, x_{k-1}, u_{k-1}),
\end{equation}
\noindent where $\xi \in \C{X}^\infty$ is a {\color{black}state-space sequence}, $u_{k} = \pi(x_{k})$, and $\gamma \in [0, 1]$ is a {\color{black}given} \emph{discount factor}.%
\footnote{According to this formulation, it is possible to evaluate $J^\pi$ on a {\color{black}state-space sequence} that is not a trajectory (which is needed for the theoretical results presented in Section \ref{sec:main_result}); in this case, even though the value of the states are not generated following policy $\pi$, in general it is still necessary to specify $\pi$ to obtain the values of the inputs $u_k$ used for the computation of the reward $r$.
When $J^\pi$ is evaluated on a trajectory, e.g., $J^{\pi_1}(\phi^{\pi_2})$, we will only consider the case in which $\pi_1 = \pi_2$.}
To find an acceptable policy, we set the following optimization problem and solve it via reinforcement learning:
\begin{subequations}\label{eq:rl_problem_objective}
\begin{align}
    \max_{\pi}
    &\ \ J^\pi(\phi^\pi(\tilde{x}_0)),&\\ 
    \text{s.t.}
    &\ \ x_{k+1} = f(x_k, u_k), & k &\in \{0, 1, 2, \dots \},\\
    &\ \ u_k = \pi(x_{k}),      & k &\in \{ 0, 1, 2, \dots \},\\
    &\ \ x_0 = \tilde{x}_0 \in \C{X}.&
\end{align}
\end{subequations}

Thus, the problem we aim to solve can be stated as follows.

\begin{problem}\label{prob:reward_shaping}
    Shape the reward function $r$ so that:
    \begin{enumerate}[(i)]
        \item\label{ite:assess_acceptable_through_high-return}
        it is possible to determine that a trajectory $\phi^\pi(\tilde{x}_0)$ is acceptable by assessing the value of     $J^\pi(\phi^\pi(\tilde{x}_0))$;
        \item\label{ite:optimal_policy_is_acceptable}
        an acceptable policy from $\tilde{x}_0$ (provided it exists) can be found by solving \eqref{eq:rl_problem_objective}.
    \end{enumerate}    
\end{problem}

\section{Main results}
\label{sec:main_result}

\begin{figure*}
    \centering
    \includegraphics[max width=\textwidth]{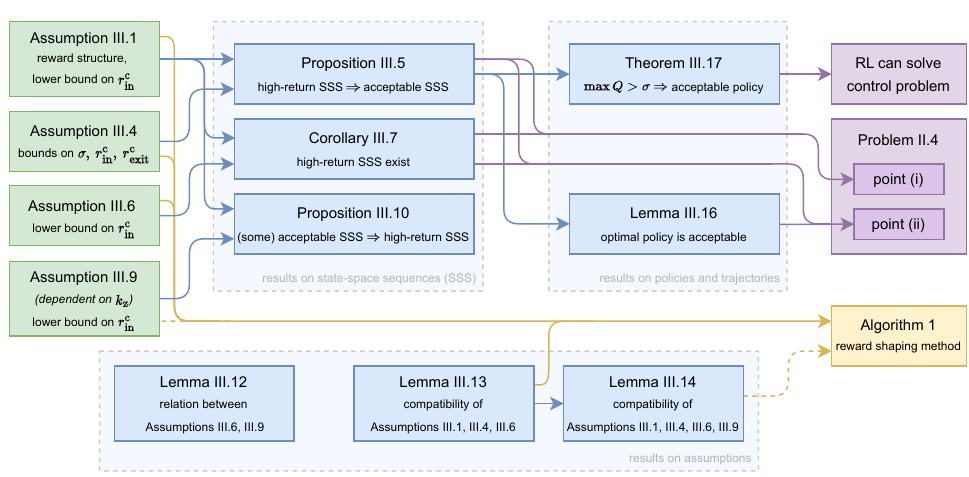}
    \caption{{\color{black}Schematic representation of the main assumptions and results in Section \ref{sec:main_result}. 
    Green blocks denote assumptions, blue blocks indicate analytical findings, yellow blocks denote algorithms, and purple blocks refer to the problems being studied.
    Dashed arrows denote optional steps in the control design. 
    ``SSS'' means ``state-space sequence''; The symbols in the figure are defined in Section \ref{sec:main_result}.}}
    \label{fig:analytical_results} 
\end{figure*}

In Section \ref{sec:assessing_acceptable_state_sequences}{\color{black},} we relate acceptable {\color{black}state-space sequences} and their return (solving point \ref{ite:assess_acceptable_through_high-return} in Problem \ref{prob:reward_shaping}),
in Section \ref{sec:algorithm_reward_shaping}{\color{black},} we embed the theory in a constructive procedure to shape rewards, 
in Section \ref{sec:assessing_acceptable_trajectories}{\color{black},} we give analogous results for trajectories, and finally in Section \ref{sec:assessing_acceptable_policies}{\color{black},} we show that acceptable policies can be found using reinforcement learning algorithms.
The assumptions we make and how they are related are schematically summarized in Figure \ref{fig:analytical_results}.

%------------------------------------
\subsection{Assessing acceptable {\color{black}state-space sequences}}
\label{sec:assessing_acceptable_state_sequences}

We start by defining the structure of the shaped reward.
\begin{assumption}[Reward structure]\label{ass:reward_structure}
The reward function can be written as
\begin{equation}\label{eq:reward_structure}
    r(x', x, u) = r^\R{b}(x', x, u) + r^\R{c}(x', x),
\end{equation}
\noindent where 
\begin{itemize}
    \item 
    $r^\R{b} :  \C{X} \times \C{X} \times \C{U} \rightarrow \BB{R}$ is a \emph{bounded reward term}, i.e., such that there exist finite $U_{\R{out}}, U_{\R{in}}, L_{\R{out}}, L_{\R{in}} \in \BB{R}$ such that
    \begin{subequations}\label{eq:bounds_reward}
    \begin{align}
        \label{eq:bound_max_reward_out}
        \sup_{x' \in \C{X} \setminus \C{G}, \ x \in \C{X}, \ u \in \C{U}}
        r^\R{b}(x', x, u) &\le U_{\R{out}},\\
        \label{eq:bound_max_reward_in}
        \sup_{x' \in \C{G}, \ x \in \C{X}, \ u \in \C{U}}
        r^\R{b}(x', x, u) &\le U_{\R{in}},\\
        \label{eq:bound_min_reward_out}
        \inf_{x' \in \C{X} \setminus \C{G}, \ x \in \C{X}, \ u \in \C{U}}
        r^\R{b}(x', x, u) &\ge L_\R{out},\\
        \label{eq:bound_min_reward_in}
        \inf_{x' \in \C{G}, \ x \in \C{X}, \ u \in \C{U}}
        r^\R{b}(x', x, u) &\ge L_\R{in}.
    \end{align}
    \end{subequations}
    
    \item
    $r^\R{c} : \C{X} \times \C{X} \rightarrow \BB{R}$ is a \emph{correction term} given by
    \begin{equation} \label{eq:correction_reward}
        r^\R{c}(x', x) = \begin{dcases}
        r^\R{c}_\R{in}, & \text{if} \ x' \in \C{G},\\
        r^\R{c}_\R{exit}, & \text{if} \ x \in \C{G}  \ \text{and} \ x' \not\in \C{G},\\
        0,   & \text{otherwise},
        \end{dcases}
    \end{equation}
    with $r^\R{c}_\R{in}, r^\R{c}_\R{exit} \in \BB{R}$.
    %with $r^\R{c}_\R{in} > 0$ and $r^\R{c}_\R{exit} < 0$. 
    Moreover, it holds that 
    \begin{equation}\label{eq:correction_in_min}
        r^\R{c}_\R{in} \ge U_{\R{out}} - L_{\R{in}}.
    \end{equation}
\end{itemize}
\end{assumption}

In practice, $r^\R{c}_\R{in}$ will typically be a positive reward for being inside the goal region, while $r^\R{c}_\R{exit}$ will normally be a negative reward for having left the goal region---please, refer to Figure \ref{fig:reward_structure} for a diagrammatic representation.

\begin{remark}[Generality of Assumption \ref{ass:reward_structure}]
    Assumption \ref{ass:reward_structure} is not too restrictive.
    Indeed, if one wants to use a preexisting reward, it is only required 
    %that
    it is bounded (see \eqref{eq:bounds_reward}).
    It can then be shaped by adding the correction term $r^\R{c}$ to it.    
\end{remark}

We also define {\color{black}the differences}
\begin{subequations}\label{difference_reward}
\begin{align}
    \label{eq:difference_reward_in}%
    \Delta_{\R{in}} &\coloneqq U_{\R{in}} - L_{\R{in}} \ge 0,\\   
    \label{eq:difference_reward_out}%
    \Delta_{\R{out}} &\coloneqq U_{\R{out}} - L_{\R{out}} \ge 0.
\end{align}
\end{subequations}

{\color{black}To assess properties of {\color{black}state-space sequences}, trajectories, and policies from their associated return, we {\color{black}define the \emph{return threshold} $\sigma \in \BB{R}$ and} introduce the following definition.

\begin{definition}[High-return {\color{black}state-space sequences}, trajectories and policies]%
\label{def:high_reward}
    A {\color{black}state-space sequence} $\xi$ is \emph{high-return} if $J^\pi(\xi) > \sigma$ for any policy $\pi$.
    A trajectory $\phi^\pi(\tilde{x}_0)$ is {high-return} if $J^\pi(\phi^\pi(\tilde{x}_0)) > \sigma$.
    A policy $\pi$ is \emph{high-return from} $\tilde{x}_0$ if $\phi^\pi(\tilde{x}_0)$ is high-return.
\end{definition}

Of the quantities introduced so far, those that we assume to be given (i.e., fixed) are $\C{G}$, $k_\R{s}$, $k_\R{p}$, $\gamma$, 
$U_\R{in}$, $U_\R{out}$, $L_\R{in}$, $L_\R{out}$; conversely, the quantities to be designed are $\sigma$, $r^\R{c}_\R{in}$, $r^\R{c}_\R{exit}$.

Next, we introduce an assumption on the correction terms in the reward.

\begin{assumption}\label{ass:inequalities_reward}
{\color{black}Assume that}
%Let $\sigma \in \BB{R}$ such that
\begin{equation}\label{eq:sigma}
    \sigma \geq \frac{U_{\R{out}}}{1-\gamma},
\end{equation}
and, given the {\color{black}desired} settling time $k_\R{s}$ and the {\color{black}desired} permanence time $k_\R{p}$, assume that 
\begin{align}
    r^\R{c}_\R{in} &\leq 
    - U_{\R{in}}
    - U_{\R{out}} \frac{1 - \gamma^{k_{\R{s}}}}{\gamma^{k_{\R{s}}}}
    + \sigma \frac{1 - \gamma}{\gamma^{k_{\R{s}}}},\label{eq:correction_in_max}\\
    r^\R{c}_\R{exit} &\leq - U_\R{out} - 
    \frac{1}{\gamma^{k_{\R{p}}-1}}
    \left[ (U_{\R{in}} + r^\R{c}_\R{in}) \frac{1 + \gamma^{k_{\R{p}}-1}(\gamma - 1)}{1-\gamma} - \sigma \right].\label{eq:correction_exit_max}
\end{align}
\end{assumption}
}

In the following Proposition, we state a key result that solves point \ref{ite:assess_acceptable_through_high-return} in Problem \ref{prob:reward_shaping}.

\begin{proposition}\label{pro:high_reward_are_acceptable}
Let Assumptions \ref{ass:reward_structure} and \ref{ass:inequalities_reward} hold. 
Then, high-return {\color{black}state-space sequences} are acceptable.
\end{proposition}

\begin{proof}
We will show that, for any policy $\pi$, if {\color{black}a state-space sequence} $\xi$ is not acceptable, then {\color{black}it} is not high-return (consequently, if $\xi$ is high-return, then it is acceptable).

$\xi$ can be not acceptable if and only if one of the following three scenarios occurs (cf.~Definition~\ref{def:acceptable}):
\begin{enumerate}
    \item\label{ite:never_reaches} 
    {\color{black}$\xi$} is never in the goal region $\C{G}$;
    \item\label{ite:enters_late} 
    {\color{black}$\xi$} is in $\C{G}$ for the first time at a time later than $k_{\R{s}}$;
    \item\label{ite:exits}
    {\color{black}$\xi$} exits from $\C{G}$ at {\color{black}time} $k_{\R{exit}}(\xi) \le k_{\R{p}}$.
\end{enumerate}
We now consider the three cases one by one and show that, for any $\pi$, if any of them occurs then it must hold that $\xi$ is not high-return, i.e., $J^\pi(\xi) \leq \sigma$.

%---------------
\paragraph*{Case \ref{ite:never_reaches}}

In this case, {\color{black}the state-space sequence is never in the goal region, that is} $\forall \, k \in [0, \infty),  x_k \not\in \C{G}$.
Therefore, {\color{black}only the third case in \eqref{eq:correction_reward} is fulfilled, for all $k$, and} we obtain $r^\R{c}(x_k, x_{k-1}) = 0$ for all $k$.
For any policy $\pi$, exploiting \eqref{eq:return}, \eqref{eq:reward_structure}, \eqref{eq:bound_max_reward_out}, and \eqref{eq:sigma}, we obtain%
\footnote{Recall that, for $|\gamma| < 1$, the geometric series is
$\sum_{k=0}^{+\infty} \gamma^{k} = \frac{1}{1-\gamma}$ 
and the truncated geometric series is
$\sum_{k=0}^{n-1} \gamma^{k}=\frac{1-\gamma^{n}}{1-\gamma}$.}
\begin{equation}\label{eq:proof_step_03}
\begin{aligned}
    J^\pi(\xi) %\coloneqq \sum_{k=1}^{+\infty} \gamma^{k-1} r(x_{k}, x_{k-1}, u_{k-1})\\
    = \sum_{k=1}^{+\infty} \gamma^{k-1} r^{\R{b}}(x_{k}, x_{k-1}, u_{k-1})\\
    \leq U_{\R{out}} \sum_{k=1}^{+\infty} \gamma^{k-1}  = \frac{U_{\R{out}}}{1-\gamma} \leq \sigma.
\end{aligned}
\end{equation}
Note that, in \eqref{eq:proof_step_03} and in the rest of the proof, the dependency of $J^\pi$ on the specific policy $\pi$ is made irrelevant by using the bounds in \eqref{eq:bounds_reward}.

%----------------
\paragraph*{Case \ref{ite:enters_late}}

Defining $k_{\R{enter}} \coloneqq (\min k \ \text{s.t.} \ x_k \in \C{G})$, we have that $k_{\R{enter}} > k_{\R{s}}$.
For the sake of simplicity and without loss of generality, assume that the state is always in the region $\C{G}$ after $k_{\R{enter}}$ (i.e., $x_k \in \C{G}, \forall \, k \ge k_{\R{enter}}$).%
\footnote{The reason why we do not lose generality is that we are interested in upper bounding $J^\pi(\xi)$ with $\sigma$, and the simplifying assumption makes $J^\pi(\xi)$ the largest possible, because the smallest reward obtainable inside $\C{G}$ (i.e., $r^\R{c}_\R{in} + L_{\R{in}}$) is at least equal to the largest reward obtainable outside $\C{G}$  (i.e., $U_\R{out}$), because of \eqref{eq:correction_in_min}.}
For any policy $\pi$,
from \eqref{eq:return}, \eqref{eq:reward_structure}, and \eqref{eq:correction_reward}, we obtain
\begin{multline}\label{eq:proof_step_01}
    J^\pi(\xi) %&\coloneqq \sum_{k=1}^{+\infty} \gamma^{k-1} r(x_k, x_{k-1}, u_{k-1})\\
    = \sum_{k=1}^{k_{\R{enter}}-1} \gamma^{k-1} r^{\R{b}}(x_k, x_{k-1}, u_{k-1})\\
    + \sum_{k=k_{\R{enter}}}^{+\infty} \gamma^{k-1} [r^{\R{b}}(x_k, x_{k-1}, u_{k-1}) + r^\R{c}_\R{in}].
\end{multline}
Exploiting \eqref{eq:correction_in_min}, and recalling that $k_{\R{enter}} > k_{\R{s}}$, from \eqref{eq:proof_step_01}, we obtain
%\footnote{On the right hand side we would have obtained if the agent entered $\C{G}$ at time $k_s + 1$}
\begin{multline}\label{eq:proof_step_02}
    J^\pi(\xi) \leq \sum_{k=1}^{k_{\R{s}}} \gamma^{k-1} r^{\R{b}}(x_k, x_{k-1}, u_{k-1})\\
    + \sum_{k=k_{\R{s}}+1}^{+\infty} \gamma^{k-1} [r^{\R{b}}(x_k, x_{k-1}, u_{k-1}) + r^\R{c}_\R{in}].
\end{multline}
Then, from \eqref{eq:proof_step_02} and exploiting \eqref{eq:bounds_reward}, we obtain
\begin{align*}
    J^\pi(\xi) &\leq U_{\R{out}} \sum_{k=1}^{k_{\R{s}}} \gamma^{k-1} + 
    (U_{\R{in}} + r^\R{c}_\R{in}) \sum_{k_{\R{s}}+1}^{\infty}
    \gamma^{k-1}\\
    &= U_{\R{out}} \sum_{k=0}^{k_{\R{s}}-1} \gamma^{k} + (U_{\R{in}} + r^\R{c}_\R{in})\gamma^{k_{\R{s}}}\sum_{k=0}^{+\infty} \gamma^{k}\\
    &= U_{\R{out}} \frac{1-\gamma^{k_{\R{s}}}}{1-\gamma} + (U_{\R{in}} + r^\R{c}_\R{in})\frac{\gamma^{k_{\R{s}}}}{1-\gamma}.
\end{align*}    
Exploiting \eqref{eq:correction_in_max}, it is immediate to see that $J^\pi(\xi) \le \sigma$.

%--------------------------------
\paragraph*{Case \ref{ite:exits}}

From the definition of $k_{\R{exit}}$ (see Sec.~\ref{sec:problem_statement}), we have $x_{k_{\R{exit}(\xi)}-1} \in \C{G}$ and $x_{k_{\R{exit}(\xi)}} \not\in \C{G}$.
From \eqref{eq:correction_in_min}, the largest $J^\pi(\xi)$ is obtained when the state{\color{black}-space} sequence $\xi$ is such that $x_k\in\C{G}$, $\forall k \in [1, k_{\R{exit}}(\xi))$, $\xi$ then exits the region $\C{G}$ at $k_{\R{exit}}(\xi) = k_{\R{p}}$, and enters again at time $k_{\R{p}}+1$.
Thus, without loss of generality, we assume this is the case.
Then, we have
\begin{align*}
    J^\pi(\xi)
    &\leq \sum_{k=1}^{k_{\R{p}}-1} \gamma^{k-1} [ r^{\R{b}}(x_{k}, x_{k-1}, u_{k-1}) + r^\R{c}_\R{in}] \\
    &\phantom{\le{}} + \gamma^{k_{\R{p}}-1} (r^{\R{b}}(x_{k_p}, x_{k_p-1}, u_{k_p-1)} + r^\R{c}_\R{exit})\\
    &\phantom{\le{}} + \sum_{k=k_{\R{p}}+1}^{\infty} \gamma^{k-1} [ r^{\R{b}}(x_{k}, x_{k-1}, u_{k-1}) + r^\R{c}_\R{in}]\\
    &\leq (U_{\R{in}} + r^\R{c}_\R{in})     
    \left[ \sum_{k=0}^{k_{{\R{p}}}-2} \gamma^{k} + \gamma^{k_\R{p}} \sum_{k=0}^{\infty} \gamma^{k}
    \right] \\
    &\phantom{\le{}} + \gamma^{k_{\R{p}}-1} (U_\R{out} + r^\R{c}_\R{exit})\\
    &= (U_{\R{in}} + r^\R{c}_\R{in})\frac{1-\gamma^{k_{\R{p}}-1} + \gamma^{k_{\R{p}}}}{1-\gamma} +
    \gamma^{\R{k_{\R{p}}-1}}(U_\R{out} + r^\R{c}_\R{exit}).
\end{align*}
Exploiting \eqref{eq:correction_exit_max}, we immediately verify that $J^\pi(\xi) \le \sigma$.
\end{proof}

Notably, Proposition \ref{pro:high_reward_are_acceptable} does not guarantee the existence of any high-return {\color{black}state-space sequence}. 
The existence of the latter is instead guaranteed by Corollary \ref{cor:detectable_good_trajectories_exist} below.
%, which also gives a lower bound on $r^\R{c}_\R{in}$.

\begin{assumption}\label{ass:suff_cond_existence_acceptable_trajectories}
Let
\begin{equation}\label{eq:bound_min_correction_in}
    r^\R{c}_\R{in} > \sigma (1 - \gamma) - L_{\R{in}}.
\end{equation}    
\end{assumption}

\begin{corollary}\label{cor:detectable_good_trajectories_exist}
Let Assumption \ref{ass:reward_structure} hold.
A sufficient condition for the existence of high-return {\color{black}state-space sequences} is that Assumption 
\ref{ass:suff_cond_existence_acceptable_trajectories} holds.
Moreover, a necessary condition for the existence of high-return {\color{black}state-space sequences} is that
\begin{equation}\label{eq:necess_cond_existence_acceptable_trajectories}
    r^\R{c}_\R{in} > \sigma (1 - \gamma) - U_{\R{in}}.
\end{equation}
\end{corollary}
\begin{proof}Let $P$ be the proposition ``$\exists \xi \in \C{X}^\infty : \forall \pi,  J^\pi(\xi) > \sigma$''.

%--------------------
\paragraph*{Assumption \ref{ass:suff_cond_existence_acceptable_trajectories} $ \Rightarrow P$}

Consider a {\color{black}state-space sequence} $\xi^\diamond = (x_0, x_1, ...)$ with all $x_k \in \C{G}$.
Then, for any $\pi$, from \eqref{eq:return}, \eqref{eq:correction_reward} and \eqref{eq:correction_in_min}, it holds that $J^\pi(\xi^\diamond) \ge \frac{r^\R{c}_\R{in} + L_\R{in}}{1-\gamma}$. Exploiting Assumption \ref{ass:suff_cond_existence_acceptable_trajectories}, we derive that $J^\pi(\xi^\diamond) > \sigma$.

%--------------------
\paragraph*{\eqref{eq:necess_cond_existence_acceptable_trajectories} $ \Leftarrow P$}

To demonstrate this result, we show equivalently that $\neg$\eqref{eq:necess_cond_existence_acceptable_trajectories} $ \Rightarrow \neg P$.
Using again
\eqref{eq:return}, \eqref{eq:correction_reward} and
\eqref{eq:correction_in_min}, for any policy $\pi$, we derive that $J^\pi(\xi) \le \frac{r^\R{c}_\R{in} + U_\R{in}}{1-\gamma}$ for all the {\color{black}state-space sequences} $\xi \in \C{X}^\infty$.
If \eqref{eq:necess_cond_existence_acceptable_trajectories} does not hold, we have $J^\pi(\xi) \le \sigma, \forall \xi \in \C{X}^\infty$.
\end{proof}
We remark that although Assumption \ref{ass:suff_cond_existence_acceptable_trajectories} is not a necessary condition itself for the existence of high-return {\color{black}state-space sequences}, it implies \eqref{eq:necess_cond_existence_acceptable_trajectories} (because of \eqref{eq:difference_reward_in}), which is one.

\begin{remark}[Selection of $k_\R{p}$]\label{rem:selection_permanence_time}
    \eqref{eq:correction_exit_max} captures the only assumption that depends on $k_\R{p}$.
    Given this assumption, it is possible to observe that $k_\R{p}$ can be set to any arbitrarily large value, thus not limiting the variety of problems that can be addressed using the present theoretical framework.
\end{remark}

To summarize, we demonstrated that it is possible to check if a {\color{black}state-space sequence} is acceptable by verifying that it is high-return.
{\color{black}Conversely, there may exist acceptable state-space sequences that are not high-return, e.g., 
those that exit (and re-enter) $\C{G}$ before $k_\R{s}$, or those that enter $\C{G}$ before $k_\R{s}$ but not early enough to collect sufficient rewards to be high-return.}
{\color{black}In some cases though, it is possible to prove that acceptable state-space sequences are high-return, such as} those that enter the goal region {\color{black}not later than a} certain time instant {\color{black}($k_\R{z}$)} and never exit it, as formalized by the next Proposition.

{\color{black}
\begin{assumption}[Dependent on the choice of $k_{\R{z}}$]\label{ass:acceptable_are_high_reward}
Given some $k_{\R{z}} \in \BB{N}_{\ge 0}$, with $k_{\R{z}} \le k_{\R{s}}$, let
    \begin{equation}\label{eq:lower_bound_cin_acceptable_are_high_reward}
        r^\R{c}_\R{in} > 
        - L_{\R{in}}
        - L_{\R{out}} \frac{1 - \gamma^{k_{\R{z}}-1}}{\gamma^{k_{\R{z}}-1}}
        + \sigma \frac{1 - \gamma}{\gamma^{k_{\R{z}}-1}}.
    \end{equation}
\end{assumption}
}

\begin{proposition}\label{pro:acceptable_are_high_reward}
Let $k_{\R{z}} \in \BB{N}_{\ge 0}$ such that $k_{\R{z}} \le k_{\R{s}}$.
{\color{black}If} Assumption \ref{ass:reward_structure} and \ref{ass:acceptable_are_high_reward} hold, {\color{black}then} {\color{black}state-space sequences} that are in $\C{G}$ for the first time at time $k_{\R{z}}$ {\color{black}or earlier} and have $k_{\R{exit}} = \infty$ are high-return.
\end{proposition}
\begin{proof}
According to the hypothesis, let $\xi = (x_0, x_1, \dots)$ be a {\color{black}state-space sequence} such that $x_k \notin \C{G}$ for $k < k_\R{z}$ and $x_k \in \C{G}$ for $k \ge k_\R{z}$.
For all policies $\pi$, from \eqref{eq:return}, \eqref{eq:reward_structure} and \eqref{eq:correction_reward}, we obtain 
\begin{multline*}
    J^\pi(\xi) 
    = \sum_{k=1}^{k_{\R{z}}-1} \gamma^{k-1} r^{\R{b}}(x_k, x_{k-1}, u_{k-1})\\
    + \sum_{k=k_{\R{z}}}^{+\infty} \gamma^{k-1} [r^{\R{b}}(x_k, x_{k-1}, u_{k-1}) + r^\R{c}_\R{in}].
\end{multline*}
Exploiting \eqref{eq:bound_min_reward_in} and \eqref{eq:bound_min_reward_out} yields
\begin{align*}
    J^\pi(\xi) &\ge
    L_{\R{out}} \sum_{k=1}^{k_{\R{z}}-1} \gamma^{k-1}
    + (L_{\R{in}} + r^\R{c}_\R{in}) \sum_{k=k_{\R{z}}}^{+\infty} \gamma^{k-1}\\
    &=
    L_{\R{out}} \sum_{k=0}^{k_{\R{z}}-2} \gamma^{k}
    + (L_{\R{in}} + r^\R{c}_\R{in}) \gamma^{k_{\R{z}}-1} \sum_{k=0}^{+\infty} \gamma^{k}\\
    &=
    L_{\R{out}} \frac{1 - \gamma^{k_{\R{z}}-1}}{1-\gamma}
    + (L_{\R{in}} + r^\R{c}_\R{in}) \frac{\gamma^{k_{\R{z}}-1}}{1-\gamma}.
\end{align*}
Given \eqref{eq:lower_bound_cin_acceptable_are_high_reward}, it follows that $J^\pi(\xi) > \sigma$.
{\color{black}Moreover, say $\xi'$ a state-space sequence that is in $\C{G}$ for the first time at some time $k_\R{z}' < k_\R{z}$, i.e., with $x_k \notin \C{G}$ for $k < k_\R{z}'$ and $x_k \in \C{G}$ for $k \ge k_\R{z}'$.
For all policies $\pi$, exploiting \eqref{eq:correction_in_min} and the fact that $J^\pi(\xi) > \sigma$, we have
\begin{align*}
    J^\pi(\xi') &\ge
    L_{\R{out}} \sum_{k=1}^{k_{\R{z}}'-1} \gamma^{k-1}
    + (L_{\R{in}} + r^\R{c}_\R{in}) \sum_{k=k_{\R{z}}'}^{+\infty} \gamma^{k-1}\\
    &\ge L_{\R{out}} \sum_{k=1}^{k_{\R{z}}-1} \gamma^{k-1}
    + (L_{\R{in}} + r^\R{c}_\R{in}) \sum_{k=k_{\R{z}}}^{+\infty} \gamma^{k-1} > \sigma.
\end{align*}

}

\end{proof}

From \eqref{eq:lower_bound_cin_acceptable_are_high_reward}, we notice that the larger $r^\R{c}_\R{in}$ is, the later {\color{black}state-space sequences} are required to be in $\C{G}$ in order to be high-return.
Moreover, it is again important to remark that while {\color{black}state-space sequences} that enter $\C{G}$ within $k_\R{z}$ time steps always exist, the same is not necessarily true for trajectories: this depends on the dynamics of the system being controlled. 

\begin{remark}[Tracking]\label{rem:tracking}
    In the results in Section \ref{sec:assessing_acceptable_state_sequences}, it was never assumed that $\C{G}$ is a fixed region in the state space.
    Indeed, it is possible to carry out the same analysis by considering a time-dependent goal region $\C{G}_k$ and, for simplicity of computation, the quantities
    \begin{align*}
        U &\coloneqq \sup_{x' \in \C{X}, \ x \in \C{X}, \ u \in \C{U}} r^\R{b}(x', x, u),\\
        L &\coloneqq \inf_{x' \in \C{X}, \ x \in \C{X}, \ u \in \C{U}} r^\R{b}(x', x, u)
    \end{align*}
    rather than $U_{\R{out}}$, $U_{\R{in}}$, and $L_{\R{out}}$, $L_{\R{in}}$ should be used in \eqref{eq:bounds_reward}, respectively.    
    This reformulation can be used to address tracking control problems.
\end{remark}

{\color{black}Reviewing the findings derived so far, a shaped reward $r$ needs to satisfy Assumptions \ref{ass:reward_structure}, \ref{ass:inequalities_reward}, \ref{ass:suff_cond_existence_acceptable_trajectories} (to exploits Proposition \ref{pro:high_reward_are_acceptable} and Corollary \ref{cor:detectable_good_trajectories_exist}) and optionally Assumption \ref{ass:acceptable_are_high_reward} (with some chosen $k_\R{z}$, to exploit Proposition \ref
{pro:acceptable_are_high_reward}); see also Figure \ref{fig:analytical_results}.
Next, we characterize the relation between these assumptions and show that they can hold simultaneously.
}

{\color{black}
%--------------------------------------------
\subsection{Compatibility of the assumptions}
\label{sec:compatibility__assumptions}

First, we give a Lemma to aid the selection of $r^\R{c}_\R{in}$.
\begin{lemma}\label{lem:compatibility_lower_bonds_c_in}
    {\color{black}Given some $k_\R{z} \le k_\R{s}$}, if Assumption \ref{ass:acceptable_are_high_reward} holds, Assumption \ref{ass:suff_cond_existence_acceptable_trajectories} also holds.
\end{lemma}
\begin{proof}
    See the Appendix.
\end{proof}

We say that two or more Assumptions are \emph{compatible} if they can hold simultaneously.
%Next, we give two Lemmas that ensure compatibility of the assumptions. 
To guarantee that high-return state-space sequences are acceptable (Proposition \ref{pro:high_reward_are_acceptable}) and that such state-space sequences exist (Corollary \ref{cor:detectable_good_trajectories_exist}), we need Assumptions \ref{ass:reward_structure}, \ref{ass:inequalities_reward}, \ref{ass:suff_cond_existence_acceptable_trajectories}, whose compatibility is ensured by the next Lemma.

\begin{lemma}\label{lem:compatibility_assumptions}
    Assumptions
    \ref{ass:reward_structure}, \ref{ass:inequalities_reward}, \ref{ass:suff_cond_existence_acceptable_trajectories}
    are compatible if 
    \begin{equation}\label{eq:bound_sigma_holistic}
        \sigma > \frac{U_\R{out}}{1-\gamma} + \frac{\Delta_\R{in} \gamma^{k_\R{s}}}{(1-\gamma)(1-\gamma^{k_\R{s}})}.
    \end{equation}
\end{lemma}
\begin{proof}
    See the Appendix.
\end{proof}

To guarantee that a class of acceptable state-space sequences are high-return, we need Assumption 
\ref{ass:acceptable_are_high_reward} (Proposition \ref{pro:acceptable_are_high_reward}), whose compatibility with previous ones is ensured by the next Lemma.

\begin{lemma}\label{lem:compatibility_assumptions_with_k_z}
    Given some $k_\R{z} \le k_\R{s}$, Assumptions
    \ref{ass:reward_structure}, \ref{ass:inequalities_reward}, \ref{ass:suff_cond_existence_acceptable_trajectories},
    \ref{ass:acceptable_are_high_reward} are compatible if 
\begin{multline}\label{eq:bound_sigma_advanced_v02}
     \sigma > 
     \frac{\gamma^{k_\R{s}}}{(1-\gamma)(1-\gamma^{k_\R{s}})}\Delta_\R{in}
     + \frac{U_\R{out}}{1-\gamma}\\
     + \frac{\gamma^{k_\R{s}}(1-\gamma^{k_\R{z}-1})}{(1-\gamma)(\gamma^{k_\R{z}-1} - \gamma^{k_\R{s}})} \left( \frac{\gamma^{k_\R{s}} \Delta_\R{in}}{(1-\gamma^{k_\R{s}})} + \Delta_\R{out}\right).
\end{multline}
\end{lemma}
\begin{proof}
    See the Appendix.
\end{proof}
}

%--------------------------------
\subsection{A constructive procedure for reward shaping}
\label{sec:algorithm_reward_shaping}

In Algorithm \ref{alg:reward_shaping}, we {\color{black}propose} a constructive procedure that can be applied to shape the reward functions used in {\color{black}Section \ref{sec:main_result}}.
To provide more flexibility, the procedure takes a preexisting reward $r^{\R{b}}$ as input, bounded according to \eqref{eq:bounds_reward}. 
If no $r^{\R{b}}$ is available, it is possible to set $r^{\R{b}} = 0$.
{\color{black}As Lemma \ref{lem:compatibility_assumptions} ensures set $\C{I}$ in the algorithm is not empty, the latter always terminates successfully.}
{\color{black}Once Algorithm \ref{alg:reward_shaping} has been used to obtain a shaped reward $r$ (thus fixing $r^\R{c}_\R{in}$, $r^\R{c}_\R{in}$, $\sigma$, which remain constant), it is possible to run a reinforcement learning algorithm to learn a suitable control policy, as explained below in Section \ref{sec:assessing_acceptable_policies}.}

\begin{algorithm}[t]
\caption{Reward shaping}
\label{alg:reward_shaping}
\KwIn{A goal region $\C{G}$,
a {\color{black}desired} settling time $k_\R{s}$, and
a {\color{black}desired} permanence time $k_\R{p}$;
a bounded reward function $r^{\R{b}}$, {\color{black}and discount factor $\gamma$.}}
\KwOut{A reward function $r$ satisfying to Assumptions \ref{ass:reward_structure}, \ref{ass:inequalities_reward}, \ref{ass:suff_cond_existence_acceptable_trajectories}.}
\BlankLine
$U_{\R{out}} \leftarrow \sup_{x' \in \C{X} \setminus \C{G}, x \in \C{X}, u \in \C{U}} r^\R{b}(x', x, u)$%
\tcp*{c.f.~Eq.~\eqref{eq:bound_max_reward_out}}
$U_{\R{in}} \leftarrow \sup_{x' \in \C{G}, x \in \C{X}, u \in \C{U}} r^\R{b}(x', x, u)$%
\tcp*{c.f.~Eq.~\eqref{eq:bound_max_reward_in}}
$L_\R{in} \leftarrow \inf_{x' \in \C{G}, x \in \C{X}, u \in \C{U}} r^\R{b}(x', x, u)$%
\tcp*{c.f.~Eq.~\eqref{eq:bound_min_reward_in}}
$\sigma \leftarrow {\color{black}
\R{rand}\left(
\frac{U_\R{out}}{1-\gamma} + \frac{(U_\R{in} - L_\R{in}) \gamma^{k_\R{s}}}{(1-\gamma)(1-\gamma^{k_\R{s}})},\  \infty \right)}$%
\label{line:selection_sigma}%
\tcp*{from Lemma \ref{lem:compatibility_assumptions}}
$\C{I} \leftarrow  \left(
    {\color{black}\sigma (1 - \gamma) - L_{\R{in}}},
    \
    - U_{\R{in}}
    - U_{\R{out}} \frac{1 - \gamma^{k_{\R{s}}}}{\gamma^{k_{\R{s}}}}
    + \sigma \frac{1 - \gamma}{\gamma^{k_{\R{s}}}}
\right]$\label{line:interval_c_in}% 
\tcp*{c.f. Eqs.~\eqref{eq:correction_in_min}, \eqref{eq:correction_in_max}, \eqref{eq:bound_min_correction_in}}
{\color{black}$r^\R{c}_\R{in} \leftarrow \R{rand}(\C{I})$\;}
$r^\R{c}_\R{exit} \leftarrow \R{rand} \Big( 
    -\infty,$ \newline
    \wrapalgo $ - U_\R{out} - \frac{1}{\gamma^{k_{\R{p}}-1}}
    \left[ (U_{\R{in}} + r^\R{c}_\R{in}) \frac{1 + \gamma^{k_{\R{p}}-1}(\gamma - 1)}{1-\gamma} - \sigma \right]
\Big)$%
\tcp*{c.f.~Eq.~\eqref{eq:correction_exit_max}}
build $r^\R{c}$ as in \eqref{eq:correction_reward}\;
$r \leftarrow r^{\R{b}} + c$\;
\end{algorithm}

It is to be noted that {\color{black}in some cases} the values of $r^\R{c}_\R{in}$ and $r^\R{c}_\R{exit}$ resulting from Algorithm \ref{alg:reward_shaping} might be significantly larger in absolute value when compared to those in $r^\R{b}$.
This can lead to a relatively \emph{sparse} reward function $r${\color{black}, i.e., one where relatively large values (in absolute value) are present but infrequent in the state-action space.
Notoriously, this lack of frequent feedback information can} make learning more difficult, {\color{black}especially} when deep reinforcement learning algorithms are used; {\color{black}see, e.g., \cite{riedmiller2018learning,rengarajan2022reinforcement} and references therein}. 
{\color{black}To mitigate this issue, it is possible to}
select $r^\R{c}_\R{in}$ and $r^\R{c}_\R{exit}$ as small in absolute value as possible, while still complying with Assumptions \ref{ass:reward_structure}, \ref{ass:inequalities_reward}, \ref{ass:suff_cond_existence_acceptable_trajectories}.
{\color{black}Reward shaping methods that do not make the reward sparse will be the subject of future work.}

\begin{remark}[Advanced reward shaping algorithm]\label{rem:advanced_algorithm}
    For simplicity, in Algorithm \ref{alg:reward_shaping}, we did not include the requirement captured by \eqref{eq:lower_bound_cin_acceptable_are_high_reward} on $r^\R{c}_\R{in}$ (used to ensure that a family of acceptable {\color{black}state-space sequences} are high-return, according to Proposition \ref{pro:acceptable_are_high_reward}), {\color{black}as} it depends on the time instant $k_\R{z}$, which would be a further parameter to select.
    This constraint can be incorporated in Algorithm \ref{alg:reward_shaping} by first selecting $k_\R{z} \le k_\R{s}$ (possibly exploiting knowledge of the system to control), {\color{black} enforcing \eqref{eq:bound_sigma_advanced_v02} at line \ref{line:selection_sigma} (Lemma \ref{lem:compatibility_assumptions_with_k_z} ensures $\C{I}$ is not empty), and using the right-hand side of   \eqref{eq:lower_bound_cin_acceptable_are_high_reward} as lower bound of $\C{I}$ at line 
    \ref{line:interval_c_in}}.
\end{remark}

%--------------------
\subsection{Assessing acceptable trajectories}
\label{sec:assessing_acceptable_trajectories}

In Section \ref{sec:assessing_acceptable_state_sequences}, we showed how the value of the return $J^\pi(\xi)$ can be used to assess whether $\xi$ is an acceptable {\color{black}state-space sequence}.
The same theory applies to trajectories (which are {\color{black}state-space sequences}, by definition).

{\color{black}It is important to remark that}, while the existence of high-return {\color{black}state-space sequences} is ensured by Proposition \ref{cor:detectable_good_trajectories_exist}, it can be much more difficult to establish if there actually exist policies that generate high-return trajectories.
This depends on the dynamics of the system at hand and the performance required, and is tightly related to the problem of reachability \cite{astrom2021feedback}, with the addition of requirements on the settling time and the permanence time.

%--------------------------------------------
\subsection{Assessing acceptable policies in value-based reinforcement learning}
\label{sec:assessing_acceptable_policies}

First, we provide a simple result stating that an acceptable policy (see Definition \ref{def:acceptable}) can be found by achieving the optimum in \eqref{eq:rl_problem_objective}, thus solving point \ref{ite:optimal_policy_is_acceptable} in {\color{black}Problem} \ref{prob:reward_shaping}.

\begin{lemma}\label{lem:optimization_problem_solves_regulation}
    {\color{black}Let Assumptions \ref{ass:reward_structure} and \ref{ass:inequalities_reward} hold.}
    If there exists a high-return policy $\pi^\diamond$ from $\tilde{x}_0 \in \C{X}$, then the optimal policy $\pi^\star$ solving the problem objective defined in \eqref{eq:rl_problem_objective} is acceptable from $\tilde{x}_0$.
\end{lemma}
\begin{proof}
    As $\pi^\star$ maximizes the return in \eqref{eq:rl_problem_objective}, then 
    $J^{\pi^\star}(\phi^{\pi^\star}(x_0)) \ge J^{\pi^\diamond}(\phi^{\pi^\diamond}(x_0)) > \sigma$, exploiting Proposition \ref{pro:high_reward_are_acceptable} {\color{black}(applicable, as Assumptions \ref{ass:reward_structure} and \ref{ass:inequalities_reward} hold)}.
\end{proof}

Proposition \ref{pro:high_reward_are_acceptable} allows to detect acceptable {\color{black}state-space sequences} by evaluating their return $J^{\pi}$. 
However, this is not normally known in a reinforcement learning setting, but it is instead approximated through a value function.
In particular, let $Q : \C{X} \times \C{U} \rightarrow \BB{R}$ be  the state-action value function associated to the \emph{greedy policy}
\begin{equation}\label{eq:greedy_policy}
    \pi_{\R{g}}(x) = \arg \max\limits_{u\in \C{U}} Q(x, u).
\end{equation}
$Q$ is normally updated iteratively with the Bellman operator so that it converges to the value of $J^{\pi_\R{g}}$, in the sense that $Q(x, u) \approx r(f(x,u), x, u) + \gamma J^{\pi_\R{g}}(\phi^{\pi_{\R{g}}}(f(x, u))$ \cite[{\color{black}Sec.}~3]{Sutton2018}.

In the next Theorem, we conclude the analysis by showing how the acceptability of a policy can be evaluated by assessing the value of $Q$.

\begin{theorem}\label{thm:convergence_greedy_policy}
Consider a state $x_k \in \C{X}$ at time $k$%
%, and let $x_{k+1} = f(x_k, \pi_{\R{g}}(x_k))$
.
Let Assumptions \ref{ass:reward_structure} and \ref{ass:inequalities_reward} hold and assume that
\begin{equation}\label{eq:Q_approximates_cost_greedy}
        \R{sign} \left(\max \limits_{u \in \C{U}} Q(x_k, u) - \sigma\right) = 
        \R{sign} \left( J^{\pi_{\R{g}}}(\phi^{\pi_\R{g}}(x_k)) - \sigma \right).
    \end{equation}
% \end{enumerate}
If $\max_{u\in \C{U}} Q(x_k, u) > \sigma$, then $\pi_\R{g}$ is an acceptable policy from $x_k$.
\end{theorem}
\begin{proof}
Exploiting \eqref{eq:Q_approximates_cost_greedy},  $\max_{u \in \C{U}} Q(x_k, u)
> \sigma $ implies that $J^{\pi_\R{g}}(\phi^{\pi_\R{g}}(x_k)) > \sigma$.
Thus, it is immediate to apply Proposition \ref{pro:high_reward_are_acceptable} (using Assumptions {\color{black}\ref{ass:reward_structure}} and \ref{ass:inequalities_reward}) to obtain that $\phi^{\pi_\R{g}}(x_k)$ is acceptable.
\end{proof}

It is important to clarify that $\phi^{\pi_\R{g}}(x_k)$ being an acceptable trajectory means that, by following policy $\pi_{\R{g}}$, 
(i) the state{\color{black}-space} sequence will be in $\C{G}$ before $k_{\R{s}}$ time instants have passed (i.e., $\exists \, k' \in [k, k + k_{\R{s}}] : x_{k'} \in \C{G}$),
and (ii) the state will not exit from $\C{G}$ before $k_{\R{p}} + 1$ time instants have passed, (i.e., $\nexists \, k'' \in [k+1, k + k_{\R{p}}] : x_{k''-1} \in \C{G}, x_{k''} \not\in \C{G}$).
Moreover, we note that \eqref{eq:Q_approximates_cost_greedy} is satisfied if $Q$ is well approximating $J^{\pi_\R{g}}$, 
in the sense that 
%$Q(x_k, \pi_{\R{g}}(x_k)) = \max_{u \in \C{U}} Q(x_k, u) \approx J^{\pi_\R{g}}(\phi^{\pi_\R{g}}(x_k))$.
{\color{black}
\begin{equation}\label{eq:condition_convergence}
    \abs{\max \limits_{u \in \C{U}} Q(x_k, u) -  J^{\pi_{\R{g}}}(\phi^{\pi_\R{g}}(x_k))}
    <
    \abs{  J^{\pi_{\R{g}}}(\phi^{\pi_\R{g}}(x_k)) - \sigma }.
\end{equation}
Indeed, \eqref{eq:condition_convergence} implies \eqref{eq:Q_approximates_cost_greedy} through Lemma \ref{lem:same_sign} in the Appendix.%
\footnote{\color{black}Assuming $J^{\pi_{\R{g}}}(\phi^{\pi_\R{g}}(x_k)) \ne \sigma$.
The case that $J^{\pi_{\R{g}}}(\phi^{\pi_\R{g}}(x_k)) = \sigma$ is however not of interest, as the trajectory $\phi^{\pi_\R{g}}(x_k)$ would not be high-return.}
\eqref{eq:condition_convergence} is fulfilled after a finite number of iterations} 
%This happens 
if the algorithm used to update the value of $Q$ is converging {\color{black}asymptotically to $J^{\pi_\R{g}}$}, which has been proved formally for reinforcement learning algorithms like SARSA and Q-learning \cite{Sutton2018}. 
{\color{black}In} the latter, the greedy policy and the function $Q$ are guaranteed to converge to the optimal policy and to its discounted return $J$, respectively{\color{black}; hence, if high-return policies exist, Lemma \ref{lem:optimization_problem_solves_regulation} guarantees that the learned policy is acceptable}.

%------------------------------------
\section{Numerical results}%
\label{sec:numerical_results}

We validate the theory presented in Section \ref{sec:main_result} by means of two representative case studies (and corresponding reinforcement learning environments, from OpenAI Gym \cite{brockman2016openai}): Inverted Pendulum \cite{GYM_pendulum} and Lunar Lander \cite{GYM_lander}.
The former is a classic nonlinear benchmark problem in control theory, whereas the latter is a more sophisticated control problem with multiple input and outputs.
In particular, we first validate Theorem \ref{thm:convergence_greedy_policy} using Q-learning to learn a policy that stabilizes an inverted pendulum within a predefined time; then, we show that the theory also holds when using a deep reinforcement learning algorithm, such as {\color{black}Double} DQN, to learn a policy able to land a spacecraft fulfilling desired time constraints.

In each scenario, the learning phase and deployment phase are repeated in $S \in \BB{N}_{>0}$ independent sessions, which are composed of $E \in \BB{N}_{>0}$
episodes.
Each episode is a simulation lasting $N \in \BB{N}_{>0}$ time steps.
Moreover, we {\color{black}always} use the $\epsilon$-greedy policy \cite{Sutton2018} {\color{black}during learning}.

{\color{black}For reproducibility, the code is available on GitHub \cite{code}.}

%-------------------------------------
\subsection{Inverted Pendulum}
\label{sec:inverted_pendulum}

%-----------------------------------------
\subsubsection{Description of the Inverted Pendulum environment}
\label{sec:pendulum_description}

\begin{figure}[t]
  \centering
  \subfloat[]{\includegraphics[min height = 3.6cm]{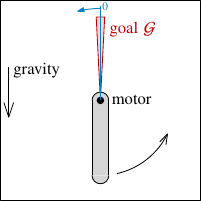}
  \label{fig:pendulum_drawing}}%
  \hfill
  \subfloat[]{\includegraphics[min height = 3.6cm]{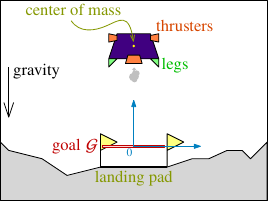}
  \label{fig:lander_drawing}}
  \caption{Sketch representation of the environments used in the numerical validation in Section \ref{sec:numerical_results}: (a) Inverted Pendulum and (b) Lunar Lander.
  Both the pendulum and the lander are depicted in their initial states.
  }
  \label{fig:environments}
\end{figure}

In this environment, the objective is to stabilize a rigid pendulum affected by gravity to the upward position in a certain time, by exploiting a torque applied at the joint.
In particular, the pendulum is a rigid rod, having length $l = 1 \, \R{m}$, mass $m = 1 \, \R{kg}$ and
moment of inertia $I = ml^2/3$; the gravitational acceleration is taken equal to $g = 10 \, \R{m/s^2}$.
A graphical depiction of the scenario is given in Figure \ref{fig:pendulum_drawing}.

%------------------------------
\paragraph{State}

The state at time $k$ is $x_k \coloneqq [x_{k,1} \ \ x_{k,2}]\T$, where $x_{k,1}$ and $x_{k,2}$ are the angular position and angular velocity of the pendulum, respectively.
In order to apply Q-learning, which is a tabular RL method, we discretize the state space (and the set of acceptable inputs). Namely, the position $x_{k, 1}$ takes values in $\left[ -{\pi}, {\pi} \right]\,\text{rad}$, with
$\left[ -\pi, -\frac{\pi}{9} \right]$ discretized into 8 equally spaced values, 
$\left( -\frac{\pi}{9}, -\frac{\pi}{36} \right]$ into 7 values, and
$\left( -\frac{\pi}{36}, 0 \right]$ into 5 values (analogously for $\left[ 0, \pi \right]$).
The velocity $x_{k, 2}$ takes values in $ \left[-8, 8 \right]\,\text{rad/s}$, with $\left[ - 8, -1 \right]$ discretized into 10 values, and
$\left( - 1, 0 \right]$ into 9 values
(analogously for $\left[ 0, 8 \right]$).
$x_{k, 1} = 0$ and $x_{k, 1} = \pi$ correspond to the upward and downward positions, respectively.
In each simulation the initial condition is chosen as $x_{0} = [\pi \ \ 0]\T$.

%------------------------------
\paragraph{Control inputs}
The control input $u_k$ is a torque applied at the pendulum's rotating joint, with values chosen in $[-2, 2]\,\text{Nm}$, with the interval $\left[ - 2, -0.2 \right]$ discretized into 9 values, 
and $\left( -0.2, 0 \right]$ into 4 values
(analogously for $\left[ 0, 2 \right]$).

%------------------------------
\paragraph{Control problem}
\label{sec:control_problem_epndulum}

Let $x^{\R{ref}} \coloneqq [0 \ \ 0]\T$ denote the unstable vertical position.
The goal region is $\C{G} \coloneqq \{x \in \C{X} \mid \norm{x - x^{\R{ref}}} < \theta\}$ {\color{black}($\norm{\cdot}$ being the Euclidean norm)}, with $\theta = 0.42$ amounting to $5\%$ of the maximum distance from the origin, in the state space.
We select the desired settling time as $k_\R{s} = 500$ time steps and the desired permanence time as $k_{\R{p}} = 1000$ time steps (cf.~Sec.~\ref{sec:problem_statement}).

%------------------------------
\paragraph{Reward}
\label{sec:reward_pendulum}

To guarantee the required performance and {\color{black}steady-state} specifications, the reward function is chosen as in \eqref{eq:reward_structure}, with $r^{\R{b}}$ being the standard Gym reward, given by
\begin{equation}\label{eq:reward_gym_pendulum}
    r^{\R{b}}(x_{k}, x_{k-1}, u_{k-1}) = - x_{1,k}^2 - 0.1 x_{2,k}^2 - 0.001 u_{k-1}^2.
\end{equation}
Following Algorithm \ref{alg:reward_shaping},
from \eqref{eq:reward_gym_pendulum}, we compute that
\begin{align*}
    U_{\R{out}} &= \max_{{\color{black}x_k \notin \C{G}}, x_{k-1} \in \C{X}, u_{k-1} \in \C{U}} r^\R{b} = - {\color{black}0.1}\theta^2 {\color{black}\approx -0.018},\\
    L_{\R{out}} &= \min_{{\color{black}x_k \notin \C{G}}, x_{k-1} \in \C{X}, u_{k-1} \in \C{U}} r^\R{b}\\
    &= - \pi^2 - 0.1 \cdot 8^2 - 0.001 \cdot 2^2 \approx -16.27, \\
    U_{\R{in}} &= \max_{{\color{black}x_k \in \C{G}}, x_{k-1} \in \C{X}, u_{k-1} \in \C{U}} r^\R{b} = 0,\\
    {\color{black}L_{\R{in}}} &=
    {\color{black}\min_{x_k \in \C{G}, x_{k-1} \in \C{X}, u_{k-1} \in \C{U}} r^\R{b} = -\theta^2 - 0.001 \cdot 2^2 \approx -0,18.}
\end{align*}
Then, {\color{black}given} $\gamma = 0.99$ [cf.~\eqref{eq:return}], {\color{black}we select} $\sigma = 10000$ [cf.~\eqref{eq:sigma}], and the correction terms in \eqref{eq:correction_reward} {\color{black}as}
\begin{align*}
    r^\R{c}_{\R{in}} &= 
        - U_{\R{in}}
        - U_{\R{out}} \frac{1 - \gamma^{k_{\R{s}}}}{\gamma^{k_{\R{s}}}}
        + \sigma \frac{1 - \gamma}{\gamma^{k_{\R{s}}}} \approx 1.52 \cdot 10^4,\\
    r^\R{c}_{\R{exit}} &=  - U_\R{out} - 
        \frac{1}{\gamma^{k_{\R{p}}-1}}
        \left[ (U_{\R{in}} + r^\R{c}_\R{in}) \frac{1+\gamma^{k_{\R{p}}-1}(\gamma -1)}{1-\gamma} - \sigma \right]\nonumber\\
        &\approx -3.50 \cdot 10^{10}.
\end{align*}

%-----------------------------------
\paragraph{Parameters}
We take $S=5$ sessions, $E=1000$ episodes, and $N=1000$ time steps.
We set the learning rate to 0.8.
For the $\epsilon$-greedy policy, we select $\epsilon = 0.05$.

%------------------------------------
\subsubsection{Results of Q-learning in the Inverted Pendulum environment}
\label{sec:results_q_learning_inverted_pendulum}

\begin{figure}[t]
\centering
\begin{minipage}{\textwidth}
\begin{tikzpicture}
    \node (img)  {\includegraphics[scale=0.5, trim = {1cm 1cm 1cm 1cm}]{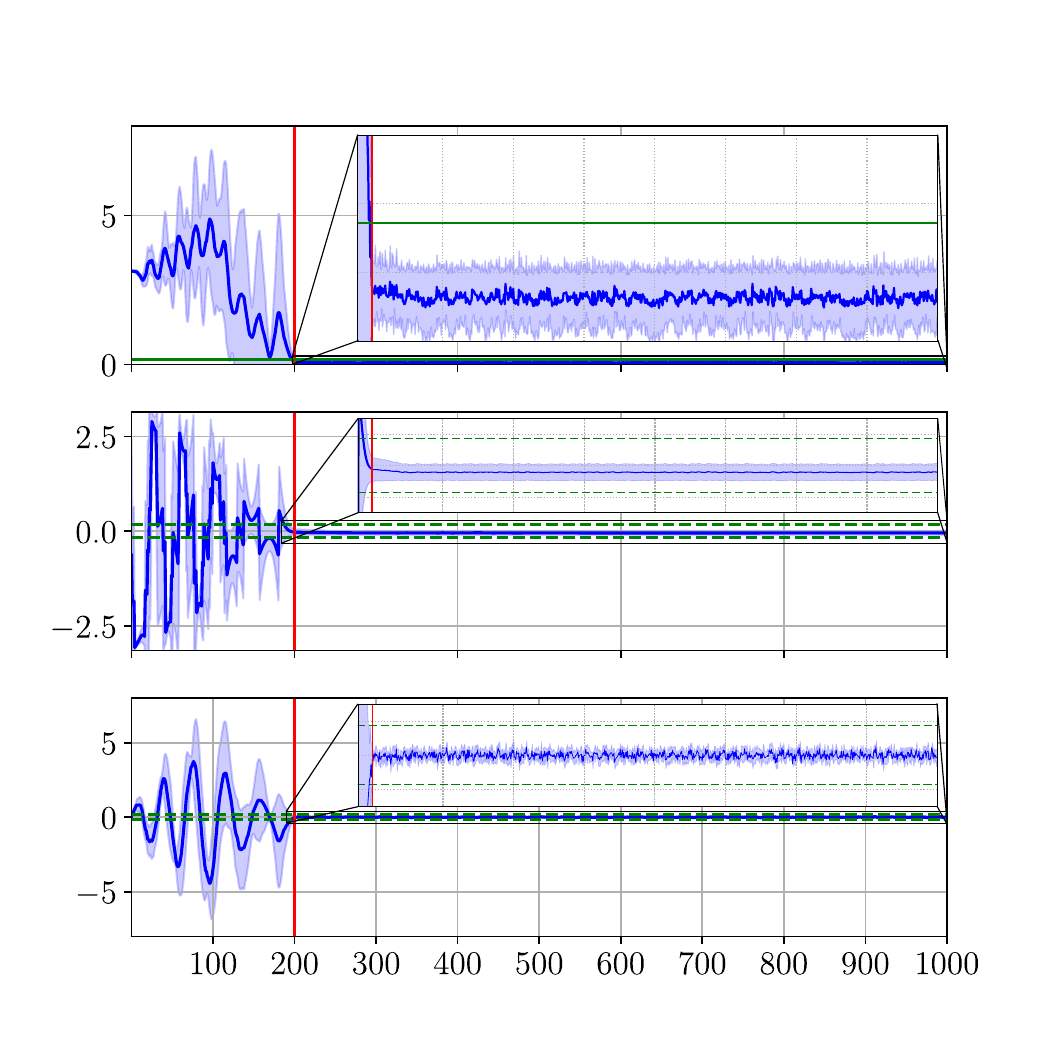}};
    \node[left=of img, node distance=0cm, rotate =90, anchor=center, yshift=-0.8cm, font=\color{black}] {Position};
    \node[left=of img, node distance=0cm, rotate =90, anchor=center, xshift=-2.5cm, yshift=-0.8cm, font=\color{black}] {Velocity};
    \node[left=of img, node distance=0cm, rotate =90, anchor=center, xshift=2.5cm, yshift=-0.8cm, font=\color{black}] {Distance from $x^\R{ref}$};
    \node[below=of img, node distance=0cm, yshift=1.1cm,font=\color{black}] {Time [steps]};
    \end{tikzpicture}
\end{minipage}\\
\caption{%
Average (blue line) plus/minus standard deviation (shaded area) of $\norm{x_k - x^\R{ref}}$ {\color{black}(top panel), angular position $x_{k,1}$ (middle panel), and angular velocity $x_{k,2}$ (bottom panel),}
obtained by $S$ policies trained with Q-learning in the pendulum environment. 
The green {\color{black}solid} line {\color{black}(top panel)} indicates the goal region {\color{black}$\C{G}$};
{\color{black}the green dashed line (middle and bottom panel) indicate neighborhoods of width $2\theta$ centered in $x_{1,k} = x^\R{ref}_1 = 0$ (middle panel) and in $x_{2,k} = x^\R{ref}_2 = 0$ (bottom panel).}
The red line indicates the time instant when the (averaged) trajectory enters the goal region.
}
\label{fig:pendulum_trajectories}
\end{figure}
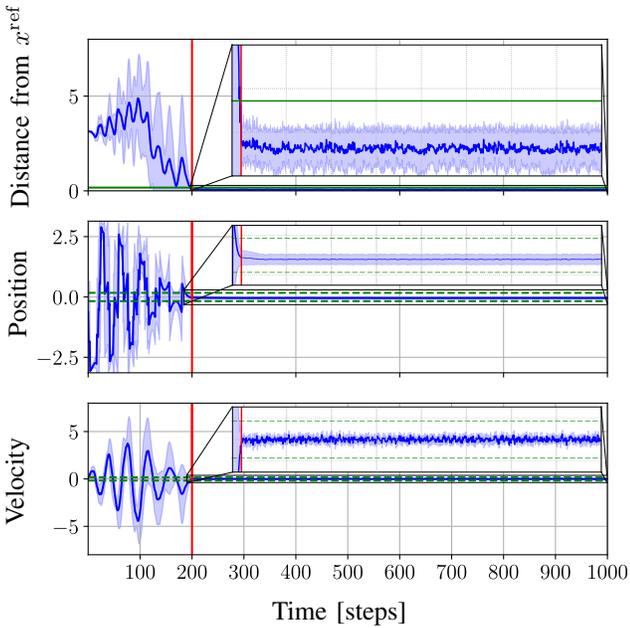

\begin{figure}[t]
\centering
\begin{minipage}{\textwidth}
\begin{tikzpicture}
    \node (img)  {\includegraphics[scale=0.5, trim = {1cm 1cm 1cm 1cm}]{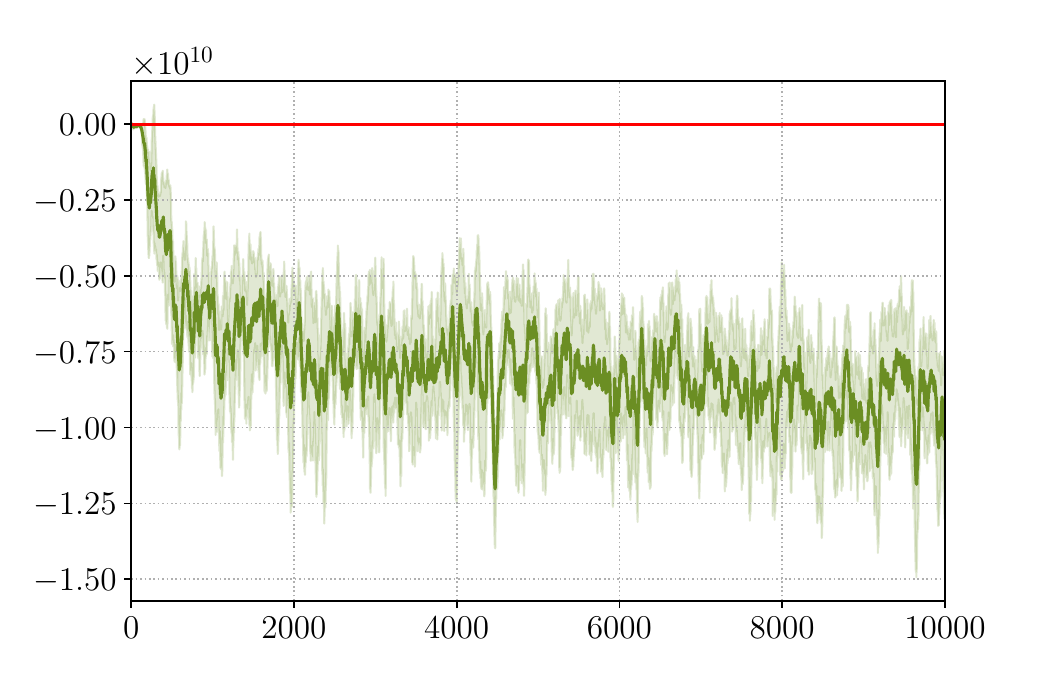}};
    \node[left=of img, node distance=0cm, rotate =90, anchor=center, yshift=-0.6cm, font=\color{black}] {Discounted return};
    \node[below=of img, node distance=0cm, yshift=0.8cm,font=\color{black}] {Episodes};
    \end{tikzpicture}
\end{minipage}
\caption{Average (green line) plus/minus standard deviation (shaded area) of the discounted returns per episode obtained in $S$ training sessions with Q-learning in the pendulum environment.
The red line indicates the threshold value $\sigma$ (cf.~Sec.~\ref{sec:assessing_acceptable_state_sequences}).
The returns are averaged backwards across episodes using a moving window of 50 samples.
}
\label{fig:pendulum_returns}
\end{figure}

After training is completed for all sessions, we test the capability of the learned policies to swing up and stabilize the pendulum within the desired settling time.
The results are portrayed in Figure \ref{fig:pendulum_trajectories}, showing {\color{black}the distance of the trajectories from $x^\R{ref}$, position, and velocity} in time.
We observe that the control problem is solved in all sessions, suggesting that the optimal policy (which would be an acceptable one, according to Lemma \ref{lem:optimization_problem_solves_regulation}) has been found.
Interestingly, this might be difficult to detect by looking only at the returns obtained during training, plotted in Figure \ref{fig:pendulum_returns}.
{\color{black}Indeed, the discounted returns per episode appear to decrease as training progresses.
This happens because, as the agent progressively learns to enter the goal region, and to do so earlier in later episodes, the chance of it incurring  the penalty $r^\R{c}_\R{exit}$ for existing the goal region increases  as the result of  random explorative actions, taken by the $\epsilon$-greedy policy used during learning. 
Although this does not prevent learning from converging to the optimal policy, in practical implementations, this can be avoided by letting the exploration rate $\epsilon$ decay in later episodes. 
However, tuning the decay rate is highly problem-dependent and no general rule can therefore be given here.

In our experiments, learning ended after the planned episodes.
An alternative heuristic method to determine when to terminate the learning stage is to pause training at regular intervals, and simulate using the greedy policy in \eqref{eq:greedy_policy}. 
If the return obtained exceeds $\sigma$, the greedy policy is deemed acceptable (see Proposition \ref{pro:high_reward_are_acceptable}), ending learning; otherwise, training continues.}

%------------------------------------
\subsection{Lunar Lander}
\label{sec:lunar_lander}

%------------------------------------
\subsubsection{Description of the Lunar Lander environment}

In a 2D space, a stylized spaceship must land with small speed in a specific area in a predetermined time, in the presence of gravity, and in the absence of friction.
The spacecraft has three thrusters to guide its descent and two supporting legs at the bottom, as depicted in Figure \ref{fig:lander_drawing}.

%-------------------------------
\paragraph{State}

The state at time $k$ is $x_k = [p_k\T \ v_k\T \ \theta_k \ \omega_k \ l_k^\R{l} \ l_k^\R{r}]\T$, where 
$p_k \in \BB{R}^2$ is the horizontal and vertical position of the center of mass (arbitrary units; a.u.), 
$v_k \in \BB{R}^2$ is its horizontal and vertical velocity (a.u./s), 
$\theta_k \in [0, 2\pi)\, \text{rad}$ is the orientation of the lander (with $0$ corresponding to the orientation of a correctly landed spacecraft),
$\omega_k$ is the rate of change of the orientation (rad/s),
$l_k^\R{l} \in \{0, 1\}$ (resp.~$l_k^\R{r}$) is $1$ if the left (resp.~right) leg is touching the ground.
The initial conditions are given by $p_0 = [0 \ 1.4]\T$ (consequently, $l_0^1 = l_0^2 = 0$), $v_0 = [0 \ 0]\T$, $\theta_0 = 0$, and $\omega_0 = 0$.
The landing area is the region $[-0.2 \ 0.2] \times [-0.001 \ 0.001]$ (horizontal and vertical intervals, respectively).
The terrain topography (beyond the landing pad) is random in each simulation.

%-------------------------------
\paragraph{Control inputs}
At each time step $k$, the lander can use at most one of its three thrusters.
In particular, we let 
$u_k^\R{m} \in \{0, 1\}$ be 1 if at time $k$ the main engine on the bottom of the spacecraft is used at full power or 0 if it is off, and define $u_k^\R{l}, u_k^\R{r} \in \{0, 1\}$ analogously for the left and right thrusters, respectively.
Then, the control input at time $k$ is the vector $u_k = [u_k^\R{m} \ u_k^\R{l} \ u_k^\R{r}]\T$, which has four possible values, depending on which thruster, if any, is used.

%-------------------------------
\paragraph{Control problem}
\label{sec:control_problem_lander}

The goal region $\C{G}$ is the set of states where $p_k$ is in the landing pad, $v_k = 0$, $\theta_k = \omega_k = 0$, and $l_k^\R{l} = l_k^\R{r} = 1$.
Additionally, we select the desired settling time as $k_\R{s} = 500$ time steps and the desired permanence time as $k_{\R{p}} = 1000$ time steps (cf.~Section \ref{sec:problem_statement}).
We also remark that the simulation stops if the lander touches the ground beyond the landing pad, {\color{black}or} if it lands on the pad with a speed that is too high.
{\color{black}During training only, the simulation is also halted if the spacecraft} lands correctly.
Further detail can be found in \cite{GYM_lander}.

%-------------------------------
\paragraph{Reward}
\label{sec:reward_lander} 

The reward function is in the form introduced in \eqref{eq:reward_structure}, with $r^\R{b}$ generated according to the standard environment definition \cite{GYM_lander}.
Namely, let $\hat{r} : \C{X} \times \C{X} \times \C{U} \rightarrow \BB{R}$ be a function given by
\begin{multline} \label{eq:hat_r}
    \hat{r}(x_k, x_{k-1}, u_{k-1}) \coloneqq 100 (\norm{p_k} - \norm{p_{k-1}})\\
    + 100(\norm{v_k} - \norm{v_{k-1}})
    + 100 (\abs{\theta_{k}} - \abs{\theta_{k-1}})\\
    + 10(l_k^\R{l} - l_{k-1}^\R{l}) + 10(l_k^\R{r} - l_{k-1}^\R{r})\\
    + 0.3 u_{k-1}^\R{m} + 0.03 u_{k-1}^\R{l} + 0.03u_{k-1}^\R{r}.
\end{multline}
Then, $r^{\R{b}}$ is given by
\begin{equation}\label{eq:reward_gym_lander}
    r^{\R{b}}(x_{k}, x_{k-1}, u_{k-1}) = \begin{cases} 
    100\phantom{-{}} \qquad    \text{if } \beta_1 \text{ is true},\\
    - 100 \qquad  \text{if } \beta_2 \text{ is true},\\
    \hat{r}\phantom{-00{}} \qquad  \text{otherwise},
    \end{cases}
\end{equation}
where $\beta_1$ and $\beta_1$ are two mutually exclusive Boolean conditions: namely, $\beta_1$ is true if the spacecraft
%comes to rest on the ground
lands on the ground and stops, and $\beta_2$ becomes true if the lander touches any point of the map with a speed that is too high (i.e., it crashes), or goes beyond the operating area of the environment, {\color{black}i.e., $[-1.5, 1.5] \times [-1.5, 1.5]$}.
Following Algorithm \ref{alg:reward_shaping},
from \eqref{eq:reward_gym_lander}, we derive that
$U_{\R{out}} = 100$,
$L_{\R{out}} = -100$, 
$U_{\R{in}} = 100$,
{\color{black}$L_{\R{in}} = 100$}.
{\color{black}Given} $\gamma = 0.99$ (cf.~\eqref{eq:return}), {\color{black}we select} $\sigma = 12000$ (cf.~\eqref{eq:sigma}), and the correction terms in \eqref{eq:correction_reward} {\color{black}as}
\begin{align*}
    r^\R{c}_\R{in} &= 
        - U_{\R{in}}
        - U_{\R{out}} \frac{1 - \gamma^{k_{\R{s}}}}{\gamma^{k_{\R{s}}}}
        + \sigma \frac{1 - \gamma}{\gamma^{k_{\R{s}}}} \approx 3.04 \cdot 10^3,\\
    r^\R{c}_\R{exit} &= -U_\R{out} - 
        \frac{1}{\gamma^{k_{\R{p}}-1}}
        \left[ (U_{\R{in}} + r^\R{c}_\R{in}) \frac{1+\gamma^{k_{\R{p}}-1}(\gamma -1)}{1-\gamma} - \sigma \right]\nonumber\\
        &\approx -6.93 \cdot 10^{9}.
\end{align*}
%-------------------------------
\paragraph{Parameters}
\label{sec:parameters_lander}

We take $S=5$ sessions, $E=1000$ episodes, and $N=1000$ time steps.
%We set the learning rate $\alpha = 0.001$.
For the $\epsilon$-greedy policy, we select $\epsilon = 0.1$.
To better stabilize the values of $Q$ and help prevent overestimation, we use {\color{black}a standard} \emph{Double DQN} {\color{black}algorithm (a variation of \emph{DQN} \cite{mnih2015human}; see \cite{doubleDQN} for a detailed description), implemented in TensorFlow 2}.
%
% \footnote{\label{fn:double_dqn}{\color{black}
% \mc{@FDL can you check please?}
% \emph{Double DQN} is a variation of the \emph{DQN} algorithm \cite{mnih2015human}.
% It uses two identical neural networks to estimate $Q$, which we might call the \emph{primary network} and the \emph{target network}.
% The primary network is updated during each episode using the standard temporal difference update rule (see, e.g., \cite[Sec. 6.5]{Sutton2018}), but the action to take is selected by maximizing $Q$ as computed by the target network, and the value of the resulting state-action pair is estimated using the primary network (c.f.~\cite[(4)]{doubleDQN}).
% {\color{teal} The primary Q-network is updated at each step of the episode leveraging value estimations of the target Q-network (c.f.~\cite[(4)]{doubleDQN}).
% On the other hand, the target network is updated at the end of each episode by copying the weights of the primary network onto the target network.}
% A detailed description of the algorithm 
% can be found in \cite{doubleDQN}; a description of the original DQN can instead be found in \cite{mnih2015human}.}}
%
For the neural networks, we used {\color{black}an input layer with 8 nodes}, 2 hidden layers each composed of 128 nodes with rectified linear unit (ReLU) activation functions, {\color{black}and an output layer with 3 nodes and linear activation functions.
% {\color{teal} Xavier uniform intialization is used to set the initial Networks' weights values \cite{glorot2010understanding}}.
The networks were trained using the Adam optimizer \cite{kingma2014adam}, with a learning rate of 0.001}.
Samples collected during training are stored in a replay buffer and at each training update a batch of 128 samples is used.

%------------------------------------
\subsubsection{Results of {\color{black}Double} DQN in the Lunar Lander environment}
\label{sec:results_lander}

\begin{figure}[t]
\centering
\begin{minipage}{\textwidth}
\begin{tikzpicture}
    \node (img)  {\includegraphics[scale=0.5,trim = {1cm 1cm 1cm 1cm}]{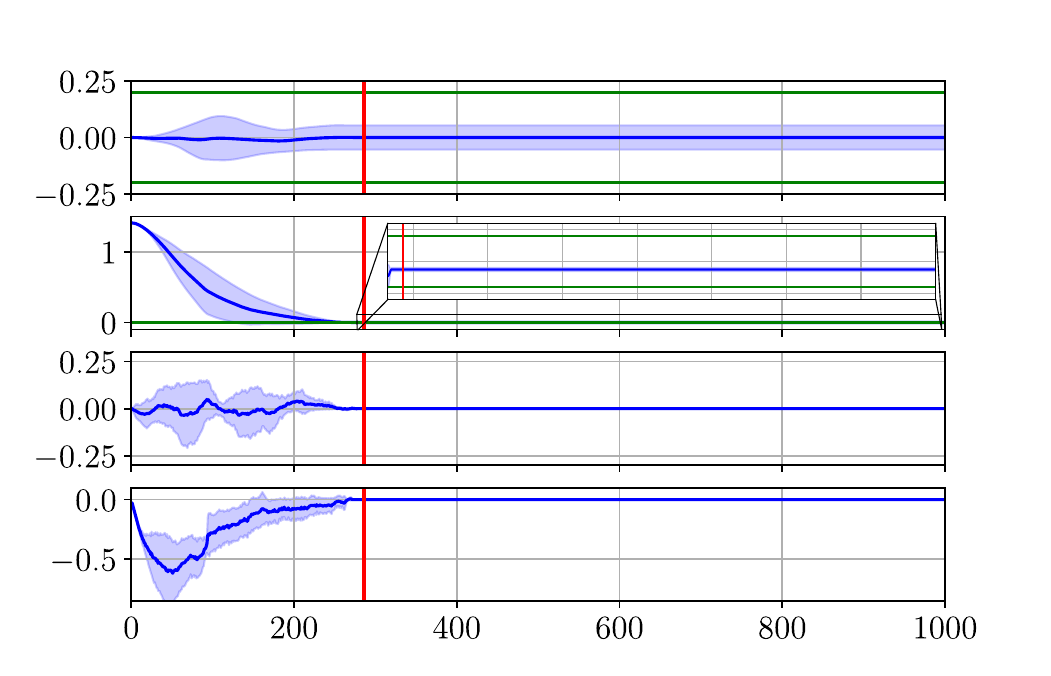}};
    \node[left=of img, node distance=0cm, rotate =90, anchor=center, yshift=-0.6cm, font=\color{black}] {Position and velocity};
    \node[below=of img, node distance=0cm, yshift=0.8cm,font=\color{black}] {Time [steps]};
    \end{tikzpicture}
\end{minipage}
\caption{Average (blue line) plus/minus standard deviation (shaded area) of the trajectory obtained by $S$ policies trained with {\color{black}Double} DQN in the Lunar Lander environment. 
{\color{black}From top to bottom: position on horizontal axis, position on vertical axis, velocity on horizontal axis, velocity on vertical axis}.
The green lines define the goal region.
The red line indicates when the (averaged) trajectory enters the goal region. 
}
\label{fig:lander_trajectories}
\end{figure}

\begin{figure}[t]
\centering
\begin{minipage}{\textwidth}
\begin{tikzpicture}
    \node (img)  {\includegraphics[scale=0.5,trim = {1cm 1cm 1cm 1cm}]{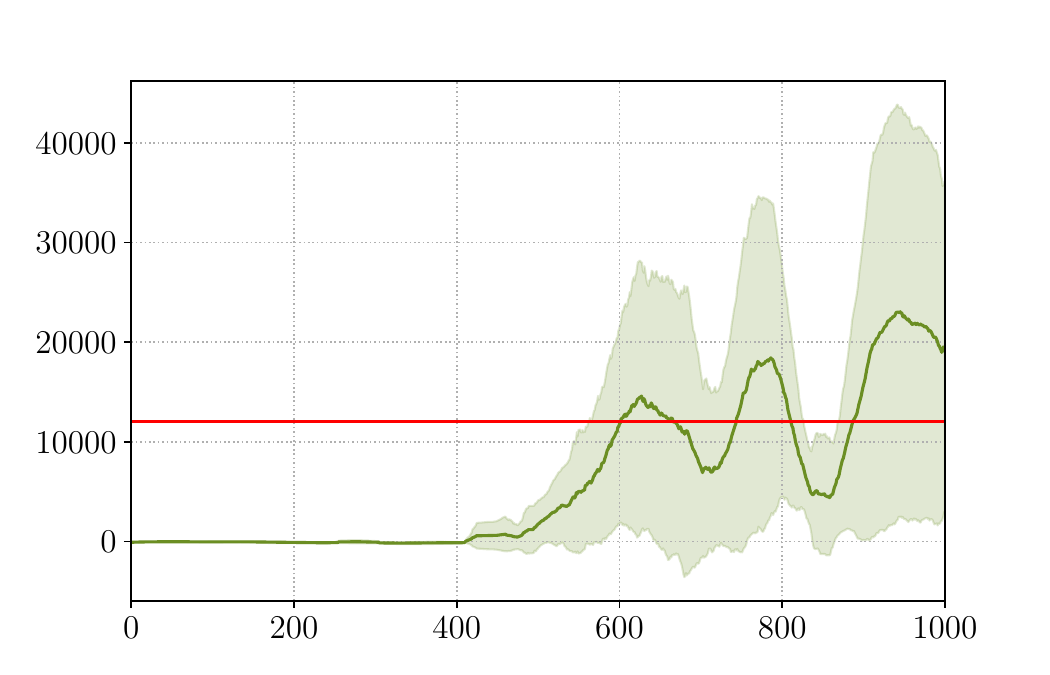}};
    \node[left=of img, node distance=0cm, rotate =90, anchor=center, yshift=-0.6cm, font=\color{black}] {Discounted return};
    \node[below=of img, node distance=0cm, yshift=0.8cm,font=\color{black}] {Episodes};
    \end{tikzpicture}
\end{minipage}
\caption{
Average (green line) plus/minus standard deviation (shaded area) of the discounted returns per episode obtained in $S$ training sessions with {\color{black}Double} DQN in the Lunar Lander environment.
The red line indicates the threshold value $\sigma$ (cf.~Sec.~\ref{sec:assessing_acceptable_state_sequences}). 
The returns are averaged backwards across episodes using a moving window of 50 samples.
}
\label{fig:lander_returns}
\end{figure}

In this environment, in all sessions, the policies learned with {\color{black}Double} DQN are able to solve the given control problem, fulfilling the given control requirements, as showed in Figure \ref{fig:lander_trajectories}.
In Figure \ref{fig:lander_returns}, we also report the returns obtained by the learning algorithm.
It is possible to observe that we obtain (averaged) returns that are over the threshold value $\sigma$.
In this case, the large negative returns, which {\color{black}were} visible in Figure \ref{fig:pendulum_returns} for the Inverted Pendulum environment, are not present.
{\color{black}The reason is that}, in the Lunar Lander simulation environment{\color{black}, during training (but not on validation)}, once the lander stops, the simulation is halted; therefore, in this case, {\color{black}during training} the lander never exits the goal region once it has entered it.

%-----------------------------------------------------
\section{Conclusions}%
\label{sec:conclusions}
One of the most significant issues holding back the use of reinforcement learning for control applications is the lack of guarantees concerning the performance of the learned policies.
In this work, we have presented analytical results that show how a specific shaping of the reward function can ensure that a control problem, such as a regulation problem, is solved with arbitrary precision, within a given settling time. 
We have validated the proposed theoretical approach on two representative experimental scenarios: the stabilization of an inverted pendulum and the landing of a simplified spacecraft.

One drawback of the present methodology is that the shaped reward might be relatively sparse {\color{black}(as discussed in Section \ref{sec:algorithm_reward_shaping})}, which could possibly hamper learning when using deep reinforcement learning algorithms.
Future work will focus on integrating existing techniques \cite{memarian2021self} (and developing new ones) for reward shaping, which are able to deal with the potential sparse reward problem, on extending the current results to the case of stochastic system dynamics, {\color{black}and on deriving conditions to ensure feasibility of a set of control requirements ($\C{G}$, $k_\R{s}$, $k_\R{p}$) for a given system, which is highly problem-dependent}.

\bibliography{support_files/references}
\bibliographystyle{IEEEtran}

%--------------------------------------
\appendix
\section{Appendix}

Let $\delta_\C{G}(x) \coloneqq \min_{y \in \C{G}} \norm{x - y}$ be the distance between $x$ and $\C{G}$;
let $h(x, u) \coloneqq f(x, u) - x$,
and let $H \coloneqq \sup_{x \in \C{X}, u \in \C{U}} \norm{g(x, u)}$.

\begin{proposition}\label{pro:settling_time_large_enough}
    Consider the system defined in \eqref{eq:dynamical_system}, and let $H$ be finite.
    If $k_\R{s} < \delta_\C{G}(\tilde{x}_0)/H$, there does not exist an acceptable policy $\pi$ from $\tilde{x}_0$.
\end{proposition}
\begin{proof}
{\color{black}We} will show that there does not exist a policy $\pi$ such that in $\phi^\pi(\tilde{x}_0)$ there exist some $k' \le k_\R{s}$ such that $x_{k'} \in \C{G}$.
Namely, consider some policy $\pi$ and the trajectory $\phi^\pi(\tilde{x}_0)$; note that
\begin{equation}\label{eq:proof_step_06}
\begin{aligned}
    \delta_\C{G}(\tilde{x}_0) &= \min_{y \in \C{G}} \norm{\tilde{x}_0 - y} =
    \min_{y \in \C{G}} \norm{\tilde{x}_0 - x_k + x_k - y}\\
    &\le 
    \min_{y \in \C{G}} \left( \norm{\tilde{x}_0 - x_k} + \norm{x_k - y} \right)\\
    &= \norm{\tilde{x}_0 - x_k} + \min_{y \in \C{G}} \norm{x_k - y}.
\end{aligned}
\end{equation} 
{\color{black}Rewrite \eqref{eq:proof_step_06} as}
\begin{equation}
    \min_{y \in \C{G}} \norm{x_k - y} = 
    \delta_\C{G}(x_k) \ge 
    \delta_\C{G}(\tilde{x}_0) - \norm{\tilde{x}_0 - x_k}.
\end{equation}
A necessary condition for obtaining $x_k \in \C{G}$ is that $\delta_\C{G}(x_k) = 0$, which is possible only if 
\begin{equation}\label{eq:proof_step_04}
    \norm{\tilde{x}_0 - x_k} \ge \delta_\C{G}(\tilde{x}_0).
\end{equation}
At the same time, it holds that
\begin{equation}
    \norm{\tilde{x}_0 - x_k} = \norm{\tilde{x}_0 - \left( \tilde{x}_0 + \sum_{j = 0}^{k-1} g(x_j, u_j) \right)} \le k H.
\end{equation}
Thus, to satisfy \eqref{eq:proof_step_04}, it is required that $kH \ge \delta_\C{G}(\tilde{x}_0)$.
Hence, if $k_\R{s} < \delta_\C{G}(\tilde{x}_0)/H$,
then surely $x_k \not\in \C{G}$ for all $k \le k_\R{s}$, and thus $\phi^\pi(\tilde{x}_0)$ is not acceptable.
\end{proof}

\begin{proof}[Proof of Lemma \ref{lem:compatibility_lower_bonds_c_in}]
We rewrite \eqref{eq:lower_bound_cin_acceptable_are_high_reward} as
\begin{equation}\label{eq:proof_step_05}
    r^\R{c}_\R{in} > - L_{\R{in}}
    + \frac{1 - \gamma^{k_{\R{z}}-1}}{\gamma^{k_{\R{z}}-1}} [(1-\gamma) \sigma - L_{\R{out}}] + (1-\gamma) \sigma.
\end{equation}
Exploiting \eqref{eq:sigma} and \eqref{eq:difference_reward_out}, we have $(1-\gamma) \sigma \ge U_{\R{out}} \ge L_{\R{out}}$, {\color{black}that is} $(1-\gamma) \sigma - L_{\R{out}} \ge 0$.
Thus, {\color{black}\eqref{eq:proof_step_05} implies \eqref{eq:bound_min_correction_in}}.
\end{proof}

{\color{black}
\begin{proof}[Proof of Lemma \ref{lem:compatibility_assumptions}]
Assumptions
\ref{ass:reward_structure}, \ref{ass:inequalities_reward}, \ref{ass:suff_cond_existence_acceptable_trajectories}
are compatible if it is possible to select the constants $\sigma$, $r^\R{c}_\R{in}$, $r^\R{c}_\R{exit}$ in accordance with 
\eqref{eq:sigma}, 
\eqref{eq:correction_in_min},
\eqref{eq:correction_in_max},
\eqref{eq:correction_exit_max},
\eqref{eq:bound_min_correction_in}.
It is always possible to select some $r^\R{c}_\R{exit}$ that satisfies to \eqref{eq:correction_exit_max}; differently, to have 
\eqref{eq:sigma},
\eqref{eq:correction_in_min}, 
\eqref{eq:correction_in_max}, \eqref{eq:bound_min_correction_in} be compatible, the following must hold:
\begin{enumerate}[(a)]
    \item\label{ite:step_proof_01}
    \eqref{eq:sigma},
    \item\label{ite:step_proof_02}
    $U_{\R{out}} - L_{\R{in}} \le - U_{\R{in}} - U_{\R{out}} \frac{1 - \gamma^{k_{\R{s}}}}{\gamma^{k_{\R{s}}}} + \sigma \frac{1 - \gamma}{\gamma^{k_{\R{s}}}}$ [from \eqref{eq:correction_in_min} and \eqref{eq:correction_in_max}].
    \item\label{ite:step_proof_03} 
    $\sigma (1-\gamma) - L_{\R{in}} < - U_{\R{in}} - U_{\R{out}} \frac{1 - \gamma^{k_{\R{s}}}}{\gamma^{k_{\R{s}}}} + \sigma \frac{1 - \gamma}{\gamma^{k_{\R{s}}}}$ [from \eqref{eq:correction_in_max} and \eqref{eq:bound_min_correction_in}].
\end{enumerate}
Note that \ref{ite:step_proof_01} holds if 
$\sigma \ge \frac{U_\R{out}}{1-\gamma}$, 
\ref{ite:step_proof_02} holds if 
$\sigma \ge \frac{U_\R{out}}{1-\gamma} + \frac{\Delta_\R{in} \gamma^{k_\R{s}}}{1-\gamma}$,
and \ref{ite:step_proof_03} holds if \eqref{eq:bound_sigma_holistic} holds, which is the most restrictive of the three.
\end{proof}

\begin{proof}[Proof of Lemma \ref{lem:compatibility_assumptions_with_k_z}]
From \eqref{eq:correction_in_max} and \eqref{eq:lower_bound_cin_acceptable_are_high_reward}, Assumptions \ref{ass:acceptable_are_high_reward} and \ref{ass:inequalities_reward} are compatible if
\begin{multline}\label{eq:proof_step_07}
     - L_{\R{in}}
        - L_{\R{out}} \frac{1 - \gamma^{k_{\R{z}}-1}}{\gamma^{k_{\R{z}}-1}}
        + \sigma \frac{1 - \gamma}{\gamma^{k_{\R{z}}-1}}\\
     < - U_{\R{in}} - U_{\R{out}} \frac{1 - \gamma^{k_{\R{s}}}}{\gamma^{k_{\R{s}}}} + \sigma \frac{1 - \gamma}{\gamma^{k_{\R{s}}}}.
\end{multline}
We will show that \eqref{eq:proof_step_07} can be rewritten as \eqref{eq:bound_sigma_advanced_v02}.
Then, recalling that $\gamma \in [0, 1]$, $\Delta_\R{in},\Delta_\R{out} \ge 0$ and $k_\R{z} \le k_\R{s}$, it is clear that \eqref{eq:bound_sigma_advanced_v02} is stricter than \eqref{eq:bound_sigma_holistic} in Lemma \ref{lem:compatibility_assumptions} (and thus implies it), hence proving the thesis that all Assumptions
\ref{ass:reward_structure}, \ref{ass:inequalities_reward}, \ref{ass:suff_cond_existence_acceptable_trajectories},
\ref{ass:acceptable_are_high_reward} are compatible.

To show that \eqref{eq:proof_step_07} can be rewritten as \eqref{eq:bound_sigma_advanced_v02}, rewrite \eqref{eq:proof_step_07} as
\begin{multline}\label{eq:proof_step_08}
     \sigma > 
     \frac{\gamma^{k_\R{s}} \gamma^{k_\R{z}-1} \Delta_\R{in}}{(1-\gamma)(\gamma^{k_\R{z}-1} - \gamma^{k_\R{s}})}
     + \frac{\gamma^{k_\R{z}-1}(1-\gamma^{k_\R{s}}) U_\R{out}}{(1-\gamma)(\gamma^{k_\R{z}-1}-\gamma^{k_\R{s}})}\\
     - \frac{\gamma^{k_\R{s}} (1-\gamma^{k_\R{z}-1}) L_\R{out}}{(1-\gamma)(\gamma^{k_\R{z}-1}-\gamma^{k_\R{s}})}.
\end{multline}
Note that $\gamma^{k_\R{z}-1}(1-\gamma^{k_\R{s}}) = (\gamma^{k_\R{z}-1} - \gamma^{k_\R{s}}) + \gamma^{k_\R{s}} (1 - \gamma^{k_\R{z}-1})$.
Hence, we have
\begin{equation*}
    \frac{\gamma^{k_\R{z}-1}(1-\gamma^{k_\R{s}}) U_\R{out}}{(1-\gamma)(\gamma^{k_\R{z}-1}-\gamma^{k_\R{s}})} = 
    \frac{U_\R{out}}{1-\gamma}
    + \frac{\gamma^{k_\R{s}}(1-\gamma^{k_\R{z}-1}) U_\R{out}}{(1-\gamma)(\gamma^{k_\R{z}-1}-\gamma^{k_\R{s}})},
\end{equation*}
and we rewrite \eqref{eq:proof_step_08} as
\begin{multline}\label{eq:proof_step_09}
     \sigma > 
     \frac{\gamma^{k_\R{s}} \gamma^{k_\R{z}-1} \Delta_\R{in}}{(1-\gamma)(\gamma^{k_\R{z}-1} - \gamma^{k_\R{s}})}
     + \frac{U_\R{out}}{1-\gamma}
     \\
     - \frac{\gamma^{k_\R{s}} (1-\gamma^{k_\R{z}-1}) \Delta_\R{out}}{(1-\gamma)(\gamma^{k_\R{z}-1}-\gamma^{k_\R{s}})}.
\end{multline}
Now, note that
\begin{multline*}
    \frac{\gamma^{k_\R{z}-1}}{\gamma^{k_\R{z}-1} - \gamma^{k_\R{s}}} = \frac{\gamma^{k_\R{z}-1}(1-\gamma^{k_\R{s}}) - (\gamma^{k_\R{z}-1}-\gamma^{k_\R{s}})}{(\gamma^{k_\R{z}-1} - \gamma^{k_\R{s}})(1-\gamma^{k_\R{s}})} \\ {} + \frac{1}{1-\gamma^{k_\R{s}}} =
    \frac{\gamma^{k_\R{s}}(1-\gamma^{k_\R{z}-1})}{(\gamma^{k_\R{z}-1} - \gamma^{k_\R{s}})(1-\gamma^{k_\R{s}})} + \frac{1}{1-\gamma^{k_\R{s}}},
\end{multline*}
and rewrite \eqref{eq:proof_step_09} as
\begin{multline*}
     \sigma > 
     \frac{\gamma^{k_\R{s}}}{(1-\gamma)(1-\gamma^{k_\R{s}})}\Delta_\R{in}
     + \frac{U_\R{out}}{1-\gamma}\\
     {} + \frac{\gamma^{2k_\R{s}}(1-\gamma^{k_\R{z}-1}) \Delta_\R{in}}{(1-\gamma)(\gamma^{k_\R{z}-1} - \gamma^{k_\R{s}})(1-\gamma^{k_\R{s}})} +
     \frac{\gamma^{k_\R{s}} (1-\gamma^{k_\R{z}-1}) \Delta_\R{out}}{(1-\gamma)(\gamma^{k_\R{z}-1}-\gamma^{k_\R{s}})},
\end{multline*}
which is immediate to rewrite as \eqref{eq:bound_sigma_advanced_v02}.
\end{proof}
}

{\color{black}
\begin{lemma}\label{lem:same_sign}
    Given two scalars $a, b \in \BB{R}$, with $b \ne 0$, if $\abs{a - b} < \abs{b}$, then $\R{sign}(a) = \R{sign}(b)$.
\end{lemma}
\begin{proof}
    We analyze separately the cases given by the combinations of the signs on $a-b$ and $b$.
    (a)
    Let $b > 0$, $a - b \ge 0$;
    we have $a \ge b > 0$, that is $a > 0$.
    (b)
    Let $b > 0$, $a - b < 0$;
    we have $-(a-b) < b$, which simplifies to $a > 0$.
    (c)
    Let $b < 0$, $a - b \ge 0$;
    we have $a - b < -b$ and thus $a < 0$.
    (d)
    Let $b < 0$, $a - b < 0$;
    we have $a < b < 0$, that is $a < 0$.
\end{proof}
}

%-----------------------------------------

\vspace{11pt}

\begin{IEEEbiography}[{\includegraphics[width=1in,height=1.25in,clip,keepaspectratio]{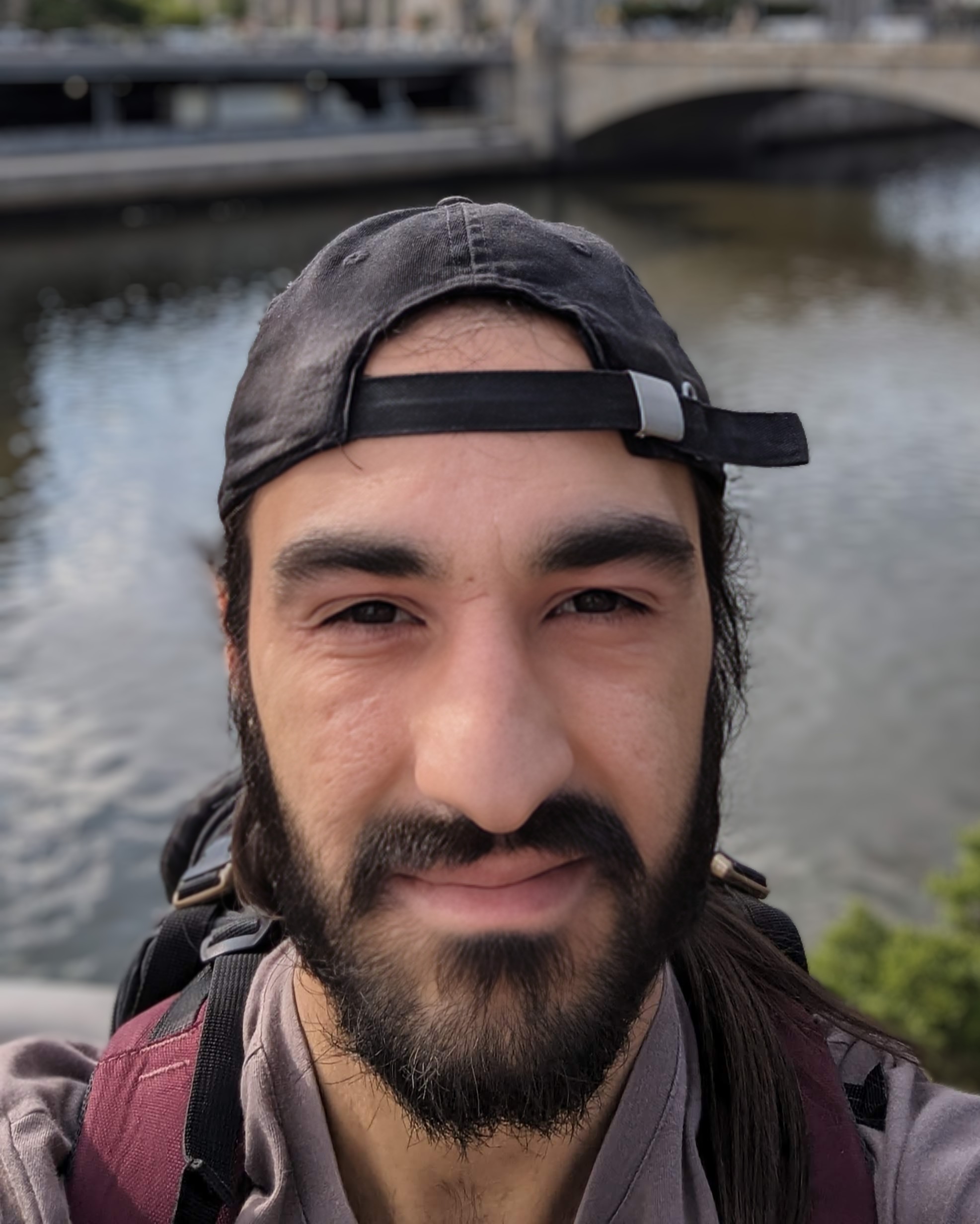}}]{Francesco De Lellis}
obtained his Ph.D. in Information Technology and Electrical Engineering at the University of Naples Federico II in May 2023. He has been a visiting researcher at the University College Dublin in 2020 and the University College London in 2022. Today, Francesco is a postdoctoral fellow with the University of Napoli Federico II.
The core of Francesco's research deals with the application of control theory, reinforcement learning and supervised learning for the development of new methodologies for the control of multi-agent complex systems.
\end{IEEEbiography}

\begin{IEEEbiography}[{\includegraphics[width=1in,height=1.25in,clip,keepaspectratio]{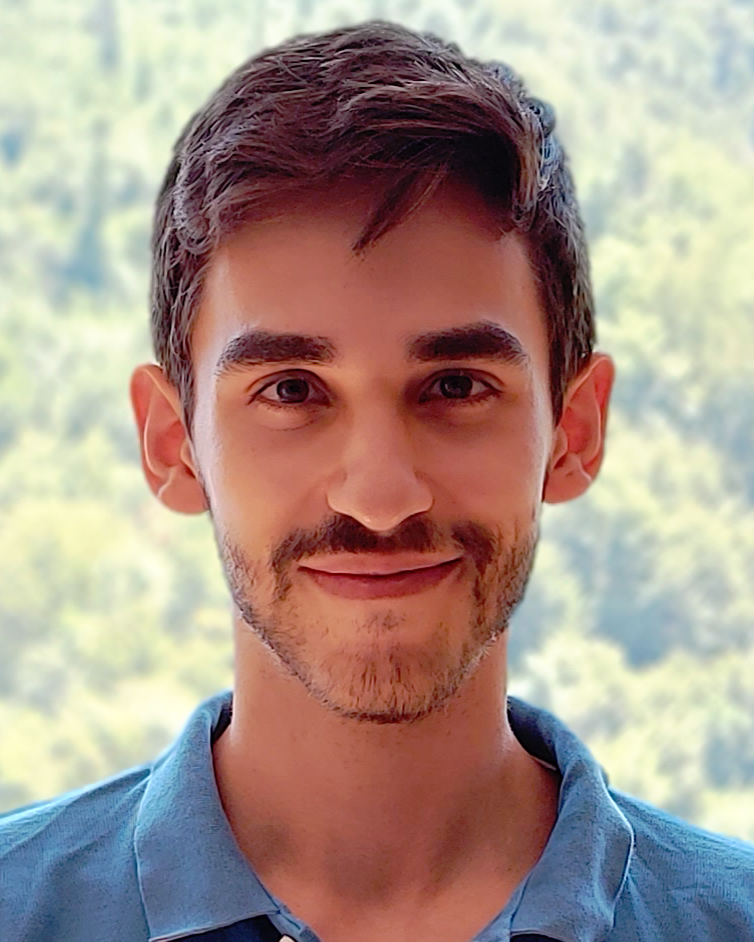}}]{Marco Coraggio}
(Member, IEEE) received the Ph.D. degree in information technology and electrical engineering from the University of Naples Federico II, Naples, Italy, in 2020.
He was a Postdoctoral Fellow with the University of Naples Federico II from 2020 to 2021 and has been a Postdoctoral Fellow with the Scuola Superiore Meridionale, School for Advanced Studies, Naples, since 2021.
He was a Visiting Student with the University of Bristol, Bristol, U.K., in 2016,
a Visiting Scholar at the University of California, Santa Barbara, CA, USA, in 2019,
and at the Link\"oping University, Link\"oping, Sweden, in 2023.
He was the finalist, in 2022, and the winner, in 2023, of the IEEE CSS Italy Young Author Best Paper Award.
His current research interests include complex networks and applications, data-driven control, and piecewise smooth and hybrid dynamical systems.
\end{IEEEbiography}

\begin{IEEEbiography}
	[{\includegraphics[width=1in,height=1.25in,clip,keepaspectratio]{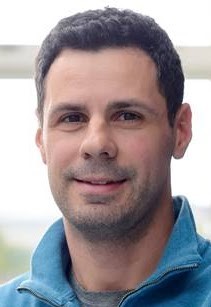}}]{Giovanni Russo}(Senior Member, IEEE) is an Associate Professor of Automatic Control at the University of Salerno, Italy. He was previously with the University of Naples Federico II (Ph.D. in 2010), Italy, Ansaldo STS (System Engineer/Integrator of the Honolulu Rail Transit Project, USA in 2012-2015), IBM Research Ireland (Research Staff Member in Optimization, Control and Decision Science from 2015 to 2018) and University College Dublin, Ireland (in 2018-2020). His research interests include contraction theory, analysis/control of nonlinear and complex systems, data-driven sequential decision-making under uncertainty and control in the space of densities. Dr. Russo has served as Associate Editor for the IEEE Transactions on Circuits and Systems I: regular papers (2016-2019) and the IEEE Transactions on Control of Network Systems (2017-2023). Since January 2024, Dr. Russo is serving as Senior Editor for the IEEE Transactions on Control of Network Systems. Personal page: \url{https://tinyurl.com/2p8zfpme}.
\end{IEEEbiography}

\begin{IEEEbiography}[{\includegraphics[width=1in,height=1.25in,clip,keepaspectratio]{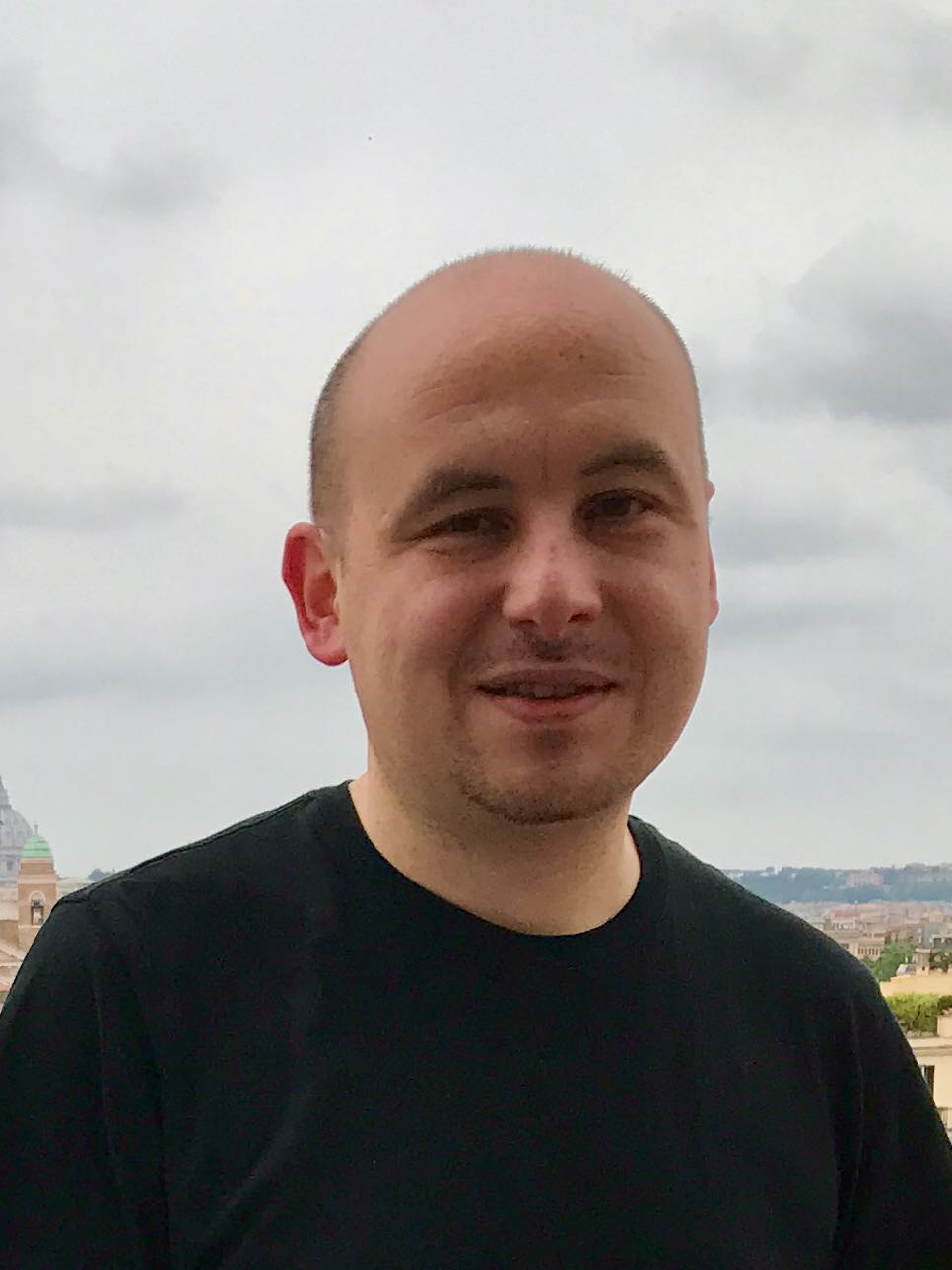}}]{Mirco Musolesi}
is Full Professor of Computer Science at the Department of Computer Science at University College London, where he leads the Machine Intelligence Lab, as part of the Autonomous Systems Research Group. He is also Full Professor of Computer Science at the University of Bologna. Previously, he held research and teaching positions at Dartmouth, Cambridge, St Andrews and Birmingham. He has broad research interests spanning several traditional and emerging areas of Computer Science and beyond.  The focus of his lab is on machine learning/artificial intelligence and their applications to a variety of theoretical and practical problems and domains. More information about his profile can be found at: \url{https://www.mircomusolesi.org}
\end{IEEEbiography}

\begin{IEEEbiography}[{\includegraphics[width=1in,height=1.25in,clip,keepaspectratio]{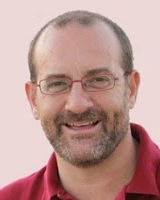}}]{Mario di Bernardo}
(Fellow, IEEE) is Professor of Automatic Control at the University of Naples Federico II, Italy and Visiting Professor of  Nonlinear Systems and Control at the University of Bristol, U.K. 
He currently serves as Deputy pro-Vice Chancellor for Internationalization at the University of Naples and coordinates the research area and PhD program on Modeling and Engineering Risk and Complexity of the Scuola Superiore Meridionale located in Naples. 
On 28th February 2007 he was bestowed the title of Cavaliere of the Order of Merit of the Italian Republic for scientific merits from the President of Italy. 
He was elevated to the grade of  Fellow of the IEEE  in January 2012 for his contributions to the analysis, control and applications of nonlinear systems and complex networks.
He was President of the Italian Society for Chaos and Complexity (2010-2017), member of the Board of Governors (2006-2011) and Vice President for Financial Activities (2011-2014) of the IEEE Circuits and Systems Society. In 2015 he was appointed to the Board of Governors of the IEEE Control Systems Society where he was elected member for the term 2023-2025. 
He was Distinguished Lecturer of the IEEE Circuits and Systems Society (2016-2017). 
He authored or co-authored more than 220 international scientific publications including more than 150 papers in scientific journals, a research monograph and two edited books.
According to the international database SCOPUS (September 2023), his h-index is 53 and his publications received over 12000 citations by other authors. 
In 2017, he received the IEEE George N. Saridis Best Transactions Paper Award for Outstanding Research. 
He was Deputy Editor-in-Chief of the IEEE Transactions on Circuits and Systems: Regular Papers, Senior Editor of the IEEE Transactions on Control of Network Systems and  Associate Editor of the IEEE Control Systems Letters, Nonlinear Analysis: Hybrid Systems, the IEEE Transactions on Circuits and Sytems I, and the IEEE Transactions on Circuits and Systems II. 
He is regularly invited as Plenary Speaker in Italy and abroad.
He has been organizer and co-organizer of several scientific initiatives and events and received funding from several funding agencies and industry including the European Union, the UK research councils the Italian Ministry of Research and University.
\end{IEEEbiography}

\vfill

\end{document}